\documentclass[sigconf]{acmart}

\renewcommand\footnotetextcopyrightpermission[1]{} 
\usepackage{tikz}
\usetikzlibrary{decorations.pathreplacing}
\usepackage{amsmath,amsfonts}
\usepackage{booktabs} 
\usepackage{multirow}
\usepackage{tabularx}

\AtBeginDocument{%
  \providecommand\BibTeX{{%
    \normalfont B\kern-0.5em{\scshape i\kern-0.25em b}\kern-0.8em\TeX}}}

\begin{document}

\title{Super Resolution Perception of Industrial Sensor Data}

\author{Jinjin Gu, Haoyu Chen, Guolong Liu, Gaoqi Liang, Xinlei Wang, Junhua Zhao}
\authornote{Junhua Zhao is the corresponding author}
\email{{jinjingu, haoyuchen, guolongliu, xinleiwang}@link.cuhk.edu.cn, {gaoqiliang, zhaojunhua}@cuhk.edu.cn}
\affiliation{%
  \institution{School of Science and Engineering, The Chinese University of Hong Kong, Shenzhen}
  \city{Shenzhen}
  \state{China}
  \postcode{518172}
}

\renewcommand{\shortauthors}{Gu, et al.}

\begin{abstract}
In this paper, we present the problem formulation and methodology framework of Super-Resolution Perception (SRP) on industrial sensor data.
Industrial intelligence relies on high-quality industrial sensor data for system control, diagnosis, fault detection, identification, and monitoring.
However, the provision of high-quality data may be expensive in some cases.
In this paper, we propose a novel machine learning problem -- the SRP problem as reconstructing high-quality data from unsatisfactory sensor data in industrial systems.
Advanced generative models are then proposed to solve the SRP problem.
This technology makes it possible to empower existing industrial facilities without upgrading existing sensors or deploying additional sensors.
We first mathematically formulate the SRP problem under the Maximum a Posteriori (MAP) estimation framework.
A case study is then presented, which performs SRP on smart meter data.
A network, namely SRPNet, is proposed to generate high-frequency load data from low-frequency data.
We further employ a novel recognition-based loss and relativistic adversarial loss to constraint the reconstruction of waveforms explicitly.
Experiments demonstrate that our SRP model can reconstruct high-frequency data effectively.
Moreover, the reconstructed high-frequency data can lead to better appliance monitoring results without changing the monitoring appliances.
\end{abstract}



\keywords{Industrial system, Industrial sensors, Signal processing, Deep learning, Super-resolution}

\maketitle

\section{Introduction}
Industry is a vital part of the economy that produces commodities in a centralized, mechanized, and automatized way.
With the introduction of Industry 4.0, a new fundamental paradigm shift in industrial production is resulted from the combination of Internet technologies and emerging technologies in the field of `smart' objects (machines and products) \cite{lasi2014industry}.
The rapid development of industrial intelligence will bring a tremendous impact on various fields of society.

Industrial intelligence is a key driver of the 4th Industrial Revolution, which can enable precise sensing and control of industrial systems through a large number of industrial sensors \cite{abramovici2015smart}.
In the modern industry, in order to improve the operating efficiency and product quality, higher accuracy of sensing and control are essential.
It will lead to a higher dependency on high precision and resolution data.

Collecting high-resolution state data of an industrial system is essential because of the following reasons.
Firstly, many industrial systems, such as aero engines, chemical processes, manufacturing systems, and power networks, are safety-critical systems.
Reliability and safety are of utmost importance to these industrial systems, which, however, are vulnerable to potential process abnormalities and component faults \cite{gao2015survey}.
Their tolerance to abnormality is very small, once small errors or abnormal fluctuations may cause serious consequences or losses.
In practice, low-resolution state data may not be sufficient for detecting or rectifying system anomalies or faults.
This may lead to permanent system damage or other potential security risks.
For instance, a $1\mathbf{Hz}$ frequency meter cannot measure the current overload that lasts less than $1\mathbf{s}$, the over-current event not detected by the low-frequency meters will lead to premature aging of the equipment and cause inestimable security risks.
The inherent defects of low-frequency meters limit the capability of an industrial system to perceive and control its internal states in a more precise manner.

Secondly, accurate state monitoring is the basis for the optimization and control of an industrial system.
In practice, the frequency of control actions is limited by the sampling frequency of meters.
If the time interval between two control actions is too large, the economic efficiency or security of the system may be compromised.
Take wind turbines as an example; the parameters of a wind turbine (e.g., blade angle) must be optimized according to the varying wind speed to increase the wind power output.
If the time interval between two wind speed measurements is $5$ minutes, it then means the wind power output of this turbine is not optimized within these $5$ minutes.
This is because the wind speed is constantly varying. The wind turbine is therefore not working in its optimal state.

In principle, installing higher-resolution meters is an ideal solution to the above problems.
However, high-resolution meters have higher costs as well.
On the other hand, this also means that a large number of low-resolution meters installed before have to be replaced, which is a huge waste of resources for large systems like power grids.
Therefore, in this paper, we propose a novel machine learning problem, namely Super Resolution Perception (SRP).
The proposed SRP problem aims at recovering high-resolution system state data from low-resolution data collected by the existing low-resolution meters, which can then be used to support more accurate and reliable state monitoring, optimization, and control.

There are at least four types of problems that can be solved by employing SRP.
Firstly, without using high-resolution meters, SRP can support the high-resolution state monitoring of industrial systems.
In practical industrial systems, sensors with different sampling frequencies may co-exist, which cannot satisfy the data quality requirement of precise control.
By converting low-resolution data to higher resolution data, SRP provides a novel and practical way to solve this problem.

Secondly, the problem of bad data and data losses can also be improved by using SRP.
It is very common that bad data are generated in some special cases during the operation of industrial systems, or the measurement data are lost during transmission.
Although state estimation and bad data detection, which are important components in industrial systems, can identify and eliminate bad data \cite{bretas2013convergence}, the removed data cannot be supplemented, which therefore influences the accuracy of estimated system statuses.
Through the detailed information provided by SRP, the bad data and missing data can not only be detected but also be well recovered, which contributes to any component that is operated based on collected data.

Thirdly, the security-constrained operation is vital for safety-critical industrial systems.
In practice, every industry system must work in its specific security region, e.g., the temperature of a boiler must be maintained within a reasonable range.
However, when the industrial sensors have insufficient sampling frequency, quality, and security variables will be difficult to measure \cite{qin2012survey}.
This will lead to the risk that the system works in an unsafe state temporarily, which will compromise the lifetime and reliability of the system.
The high-resolution data produced by SRP can help precisely control the state variables within the security regions and thus mitigate this kind of risk.
\begin{figure}[t]
    \centering
    \tikzstyle{block} = [draw,  rectangle, minimum height=2em, minimum width=4em]
    \tikzstyle{sum} = [draw, circle, node distance=1cm]
    \tikzstyle{input} = [coordinate]
    \tikzstyle{output} = [coordinate]
    \tikzstyle{pinstyle} = [pin edge={to-,thin,black}]
    \begin{tikzpicture}[auto, >=latex]
        \node [block, name = environment] (environment) {Environment};
        \node [block, right of=environment, node distance=3.5cm] (sensors) {Sensors};
        \node [block, below of=sensors, node distance = 1.5cm] (controllers) {Controllers};
        \node [block, left of=controllers, node distance=3.5cm] (actuators) {Actuators};
        \draw [->] (environment) -- (sensors);
           \draw [->] (sensors) -- (controllers);
        \draw [->](controllers) -- (actuators);
        \draw [->](actuators) -- (environment);
    \end{tikzpicture}
    \caption{Flowchart of general control process.}
    \label{fig:general_control}
    \vspace{-5mm}
\end{figure}
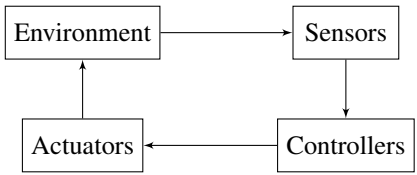

\begin{figure}[t]
    \centering
    \tikzstyle{block} = [draw,  rectangle, minimum height=2em, minimum width=4em]
    \tikzstyle{block_added} = [draw,  rectangle, dashed, minimum height=2em, minimum width=4em]
    \tikzstyle{sum} = [draw, circle, node distance=1cm]
    \tikzstyle{input} = [coordinate]
        \tikzstyle{output} = [coordinate]
    \tikzstyle{pinsty    le} = [pin edge={to-,thin,black}]

    \begin{tikzpicture}[auto, >=latex]
        \node [block, name = environment] (environment) {Environment};
        \node [block, right of=environment, node distance=2.5cm] (sensors) {Sensors};
        \node [block_added, right of=sensors, node distance=3cm, align=center] (SRP) {Super Resolution\\Perception};
        \node [block, below of=SRP, left of = SRP, node distance = 1.5cm] (controllers) {Controllers};
        \node [block, left of=controllers, node distance=2.5cm] (actuators) {Actuators};
        \draw [->] (environment) -- (sensors);
          \draw [->](sensors) -- (SRP);
           \draw [->] (SRP) |- (controllers);
        \draw [->](controllers) -- (actuators);
        \draw [->](actuators) -| (environment);
    \end{tikzpicture}
    \caption{Flowchart of control process with SRP.}
    \label{fig:with_srp}
    \vspace{-5mm}
\end{figure}
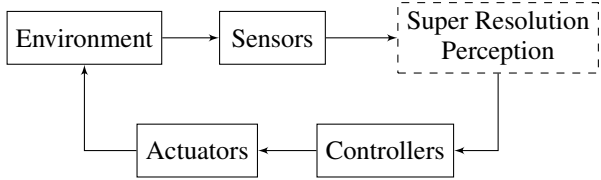

At last, high-precision control strategies require high-resolution data.
The general flowchart of a control process is shown in \figurename~\ref{fig:general_control}.
Traditional industrial control systems (ICS) are based on mechanical or electrotechnical devices and closed systems.
Today, these systems have become more expensive to deploy, maintain, and operate due to the higher functional requirements of modern industry.
To address these challenges, more communication devices and sensors are being integrated into these systems, which in turn makes the systems more complex at the hardware level \cite{kriaa2015survey}.
Different from relying on high-resolution sensors, SRP can reproduce high-resolution state data from low-resolution sensors and support the high-precision control (shown in \figurename~\ref{fig:with_srp}).
This comes with almost no increase in system complexity and additional investments in sensors and communications.

\subsection{Contributions}
In summary, this paper introduces a novel machine learning problem called Super-Resolution Perception, to recover high-resolution industrial sensor data from low-resolution observations.
Our main contributions in this paper are:
\begin{itemize}
  \item We are among the first to propose the problem of SRP for industrial sensor data. Specifically, we propose a framework based on low-frequency sampling data to have a more precise perception of industrial systems.
  \item The method of applying deep learning to solve temporal dimension SRP, namely SRPNet, is proposed for the first time. Considering the security problems, such as power overload in industrial systems, this paper proposes a new feasible method to ensure the security of the system.
  \item To reconstruct the waveforms of high-frequency data accurately, we explicitly constraint the reconstruction of waveforms by employing a novel recognition-based loss and relativistic adversarial loss. Experiments demonstrated the effectiveness of the proposed objective functions.
  \item In this paper, we prove the value and effectiveness of SRP using smart meter data as a case study. Based on the high-resolution data produced by SRP, better control strategies and control effects can be achieved for smart grids.
\end{itemize}

\subsection{Organization of the Paper}
The remainder of the paper is organized as follows.
%
%
Sec.\ref{Sec:formulation} presents the problem formulation and methodology framework of the proposed SRP on industrial sensor data.
Sec.\ref{Sec:casestudy} presents a real-world application of SRP on smart meter data.
We then explore the potential of applying SRP in Non-Intrusive Load Monitoring and show the application value of SRP. 
At last, Sec.\ref{Sec:diss} briefly discusses other potential applications of SRP and future work.

\section{Problem Formulation}
\label{Sec:formulation}
\begin{figure}
    \center
    \includegraphics[width=0.9\linewidth]{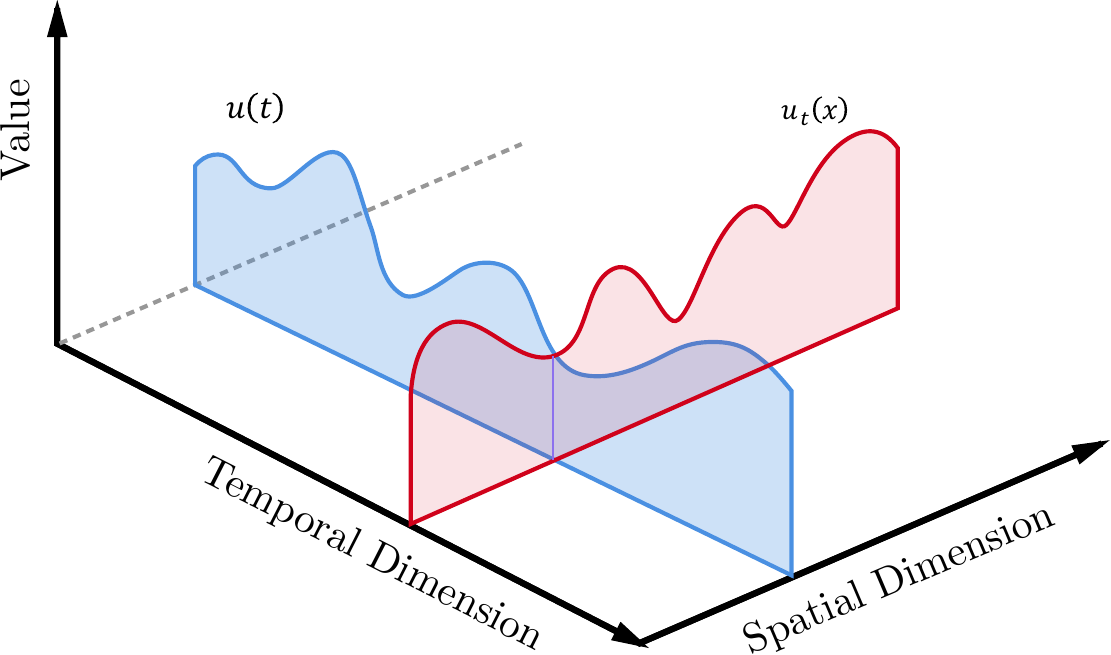}
    \caption{The relationship of temporal dimension and spatial dimension. When $u(t)$ is a scalar, u is a function of time $t$. When $u(t) = u_t(x)$ is a function of $x$ in spatial dimension, $u$ can be viewed as a multivariable function of $t$ and $x$.}
    \label{fig:ST}
    \vspace{-5mm}
\end{figure}
Consider a continuous-space physical quantity $u$.
    The value of $u$ at time $t$ can be denoted by $u(t)$.
    Let $l$ denote a low-resolution sensor data and $h$ denote the high-resolution sensor data of the same physical quantity.
    The low-res data is generated from $u$ by a continuous sampling function $\delta_L$ and the value of $l$ at time period $[t_{start}, t_{end}]$ can be wrote as
    $$l[t_n]=\int u(t)\delta_L(t-t_n)dt+n, t_n\in[t_{start}, t_{end}],$$
    where $n$ is noise.
    The high-res data $h$ of same time period but with a finer grid is as
    $$h[t_m]=\int u(t)\delta_H(t-t_n)dt+n, t_m\in [t_{start}, t_{end}],$$
    where $\delta_H$ denote high-resolution sampling function and the time index $t_m$ of $h$ is more dense.
    Note that at same time $t_0$, $h[t_0]$ and $l[t_0]$ may not equal because of the difference of $\delta_H$ and $\delta_L$.
    Both $l$ and $h$ represent the value of the same physical quantity in the same time period, but $h$ contains richer information on describing $u$.
    $h$ and $l$ are related by a degradation model:
    $$l=\downarrow h+n,$$
    where $\downarrow$ is the degradation function and $n$ is noise.
    SRP can be viewed as trying to inference $h$ with $l$ as input.
    SRP aims to find a reconstruction mapping $f$ that the reconstructed high-res data $h'=f(l)$ that recovers the information lost by degradation function as much as possible.
    
    The purpose of SRP is different depending on the property of $u$, and degradation function $\downarrow$.
    If the loss of information caused by degradation function is reflected in frequency reduction in the temporal dimension, the problem of restoring such temporal information is called \emph{Temporal SRP} problem.
    According to different application scenarios, $u(t)$ can be a scalar or a function on spatial dimension $u(t)=u_t(x)$, where $x$ denote the spatial position.
    If the degradation process loses the spatial information, the corresponding SRP problem is called \emph{Spatial SRP} problem.
    The SRP problem considering both temporal and spatial dimensions is called \emph{Spatial-Temporal SRP} problem.
    The relationship between the temporal and spatial dimension is shown by \figurename~\ref{fig:ST}.
    For different kinds of degradation functions, the SRP problem can be viewed as a superset of the various data quality problems, such as incomplete data, bad data, and malicious data, etc.

    \subsection{Interpretation of SRP as MAP estimation}


\begin{figure}[t]
    \centering
    \includegraphics[width=1.0\linewidth]{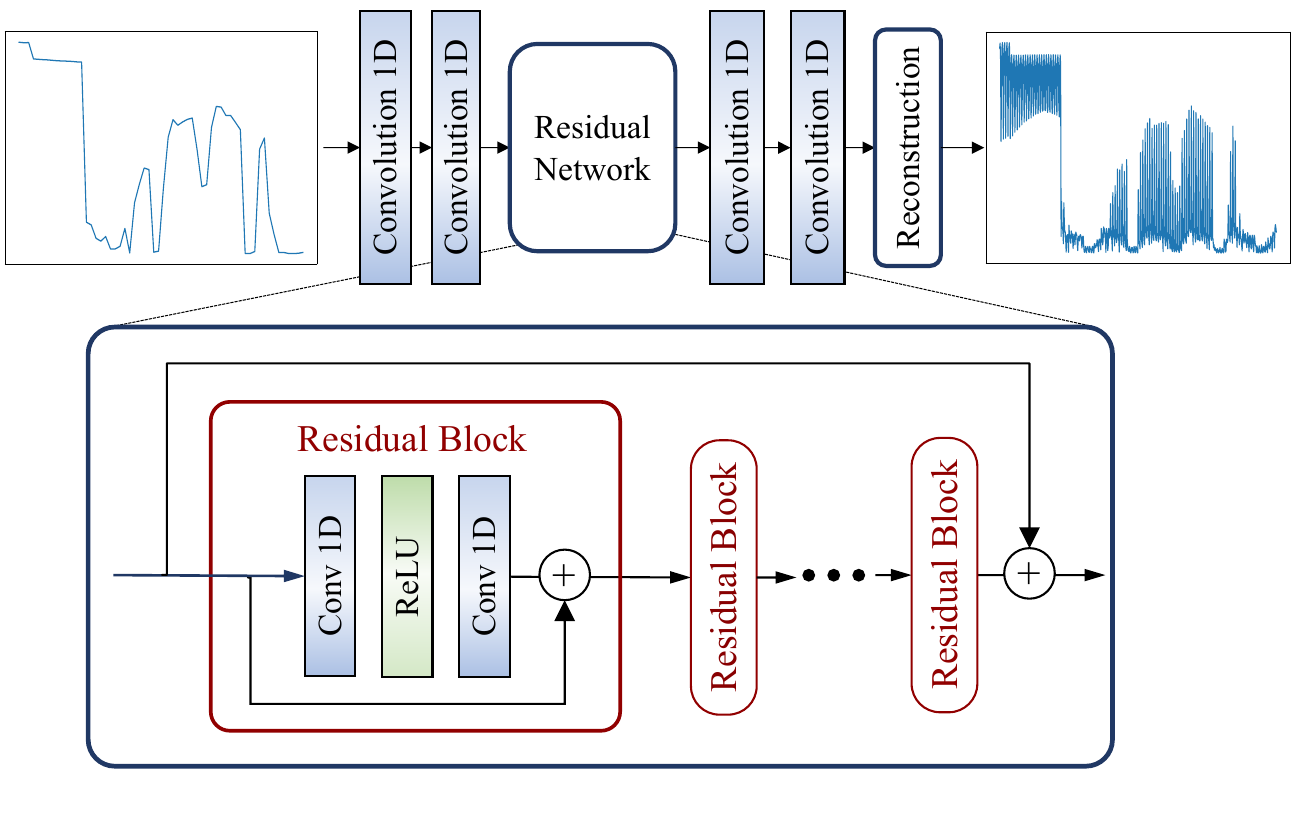}
    \vspace{-8mm}
    \caption{The proposed Super-resolution Preception Network (SRPNet). With two convolution layers for feature maps extraction, and a sub-pixel convolution layer that aggregates the feature maps from LR space and builds the SR image in a single step.}
    \label{fig:Net}
    \vspace{-5mm}
\end{figure}
    We next show that the SRP can be viewed as Maximum a Posteriori (MAP) estimation.
    For a given low-res sensor data $l$, there are many possible $h$ satisfying the degradation function.
    The final estimated $h$ is the solution with maximum posterior probability $p(h|l)$.
    According to the Bayesian formula, the posterior probability can be written as
    $$p(h|l)=\frac{p(l|h)p(h)}{p(l)},$$
    where $p(l|h)$ is the likelihood, $p(h)$ is the prior on $h$ and $p(l)$ is a constant when $l$ is given.
    The corresponding $h$ given a specific $l$ can be estimated by solving the MAP problem
    $$h'=\mathop{\arg\max}_h p(h|l)=\arg\max_h p(l|h)p(h),$$
    which is equivalent to solving the following formula
    $$h'=\mathop{\arg\max}_h \log p(l|h) + \log p(h),$$
    in which $p(l|h)$ can be solved by modeling the degradation process and $p(h)$ is obtained by solving the prior model.
    This indicates that the $h$ estimated by the SRP should not only satisfy the degradation model but also satisfy the priori characteristics of $h$.
    The prior term $p(h)$ can be viewed as a regularization term.
    Effectively modeling the prior of $h$ is important to SRP problems.
    
    This also explains how SRP reconstructs information that is not collected by low-resolution data.
    There are two situations to this problem:
    (1) The low-res data did not record the event directly.
    However, an event will last for a period of time; it is thus possible to recover this event by relying on the part of the information produced by this event and recorded in low-res data.
    (2) The information of one event is completely lost.
    In this case, it is still possible to reasonably estimate the occurrence of an event by the prior knowledge of the event as long as the event is not independent.
    If one event is independent and the information is completely lost, this is beyond the capability of the proposed SRP.

\section{Case Study}
\label{Sec:casestudy}
In this paper, we study the SRP problem using smart meter data as an example.
Smart meter data is collected by a single measurement unit (meter), and it records the overall electricity consumption of a household.
This data can be used for monitoring, analysis, and identification of load.
The frequency of the meter data depends on the sampling frequency of the smart meter.
High-frequency meter data contains more information and can be better used for monitoring and analysis of electricity consumption.
However, existing measurement units are mostly low-frequency units, which cannot produce high-frequency observation data directly.
We use SRP to process low-frequency load data and reproduce possible high-frequency data with an end-to-end neural network SRPNet.
We further prove that the use of estimated high-frequency data can lead to better appliance monitoring results without changing the monitoring method.
In this case, the physical quantity $u(t)$ represents the instantaneous power at time $t$, which is a scalar.
The load data in a certain period of time appears as a sequence. 
For a given time period, we denote the low-res load data collected by low-frequency meters as $x$ with frequency $f_l$.
The corresponding high-res data is $y$, and with a higher frequency of $f_h$.
For a single meter, $x(t)$ is a scalar and $x\in\mathbb{R}^d$, $d$ denote the length of low-res data.
With a super-resolution factor $\alpha$, we have $f_h=\alpha f_l$ and thus $y\in\mathbb{R}^{\alpha b}$.
According to the degradation model, we have $x=\downarrow y+n$.
Our goal is to recover $y$ from $x$ by a SRP mapping $f$.

\subsection{Prior Works}\label{Sec:relatedwork}
We first review the prior works of the High-Resolution Data Generation.
Due to the powerful approximation ability of deep neural networks, deep learning-based generative models have been widely used for data generation and processing. 
Deep generative models are also used in image super-resolution and audio super-resolution.
Dong et al. \cite{dong2016image} first train a three-layer convolutional neural network (CNN) for image super-resolution.
In past years, various deep learning-based methods with different network architectures \cite{VDSR,DRCN,LapSRN,MemNet,DenseSR,haris2018deep,RCAN} and training strategies \cite{ESRGAN,gu2019blind,gu2020image,feng2019suppressing} have been proposed to improve the SR performance continuously.
Advanced machine learning methods are also used to super-resolve audio data.
Dong \textit{et al.} \cite{dong2015audio} learn an analysis dictionary from the spectrogram of some related audio signals.
And then, the learned dictionary is then applied in a $l_1$-norm regularization term for the reconstruction of the high-resolution spectrogram.
Mandel \textit{et al.} \cite{mandel2015audio} utilize a non-linear dictionary-based denoising system to transform low-bandwidth, low-bitrate speech into high-bandwidth, high-quality speech.
Kuleshov \textit{et al.} \cite{kuleshov2017audio} propose an encoder-decoder network to super-resolve audio data and achieve the state-of-the-art performance in reconstructing high resolution audios.

Although generative models have been very effective in many signal processing tasks, however, to the best of our knowledge, there is no such method designed to super-resolve industrial sensor data.
Compared with audio super-resolution, the super-resolution factor of industrial sensor data are often 10, 100, or even 1000. It is much larger than that of audio, which only is 2, 5, or 10.
Industrial sensor data usually contains more distinct patterns than natural audio data.
These unique characteristics of industrial sensor data also make it possible to support large SRP factors such as 100 and 1000.

\subsection{Method}
We develop a deep convolutional neural network called SRPNet to implement the abovementioned mapping.
In this section, we first present the network architecture of SRPNet.
We next present the used objective function.
At last, we present the training details of SRPNet.
\subsubsection{Network Architecture}
\label{sec:netar}
The network architecture of SRPNet is shown in \figurename~\ref{fig:Net}.
SRPNet takes the low-frequency data as the input and outputs the estimated high-frequency data directly.
At the top of the network, two 1D convolution layers are operated as a global feature extractor.
This extractor extracts features from low-res input and represents the features as many feature vectors.
These vectors contain the abstract feature information of input, and each vector has the same dimension as input data.
After feature extraction, an information supplemental sub-network with the residual structure is used to supplement the lost information to the feature vectors.
The residual structure consists of a big global residual connection and many local residual blocks.
The global residual connection will force the network to learn the lost information rather than form the signal itself.
The local residual blocks make it possible to train a deeper network \cite{he2016deep}.
In SRPNet, we use 16 local residual blocks for better performance.
The third part is a reconstruction sub-network.
In this part, the feature vectors are integrated into $\alpha$ sub-sequences $\phi\in\mathbb{R}^{\alpha\times d}$ by two 1D convolution layers.
Then rearrange the $\alpha$ channels of the same pixel of the sub-sequences into a new sequence of length $\alpha$, corresponding to a $\alpha$ size sequence fragment in the high-res data.
Thus the $\alpha$ sub-sequences with length $d$ are rearranged into the estimated high-res sequence with length $\alpha\times d$.
This sequence is expected to be similar to the ground truth $y$.

The proposed SRPNet is mainly distinguished from the existing time series data generation methods in two aspects:
(1) Different from WaveNet \cite{van2016wavenet}, which uses a recursive strategy to generate sequence data, we use convolution operation to generate high-res sequences in parallel.
This saves unnecessary calculations and provides higher computational efficiency when generating long sequences.
For WaveNet, it takes about 60 minutes to generate 10000 samples.
And it only takes 0.2 seconds for SRPNet to generate a sequence with the same length.
(2) In SRPNet, the feature extraction and information supplement are performed on low-res sequences, and the upsampling function is implicitly included in the previous convolutional layers, which can be learned automatically.
This not only reduces the computation complexity but also makes it easier to generate the desired waveform when SRP.
Compared with AudioNet \cite{kuleshov2017audio}, which first upsample the sequences and then change the waveform with a neural network, SRPNet is able to perform SRP with higher SR factors, shorter running time, and better SR performance.

\subsubsection{Objective Function}
\label{sec:object}

\begin{figure*}[!htb]
    \centering
    \begin{tabular}{cccc}
        \includegraphics[width=.225\textwidth]{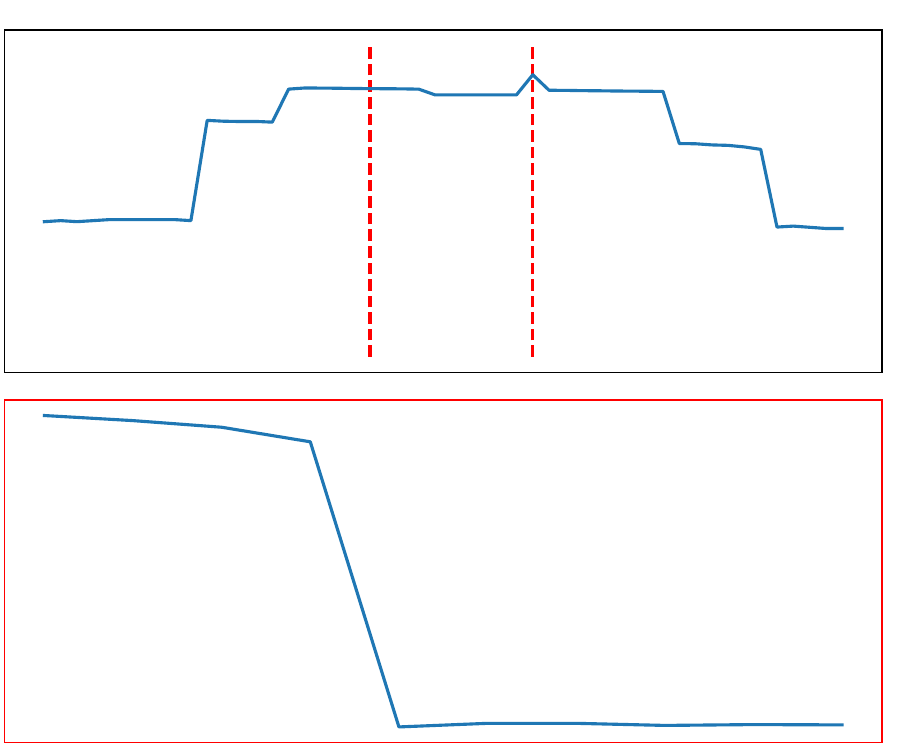} &
        \includegraphics[width=.225\textwidth]{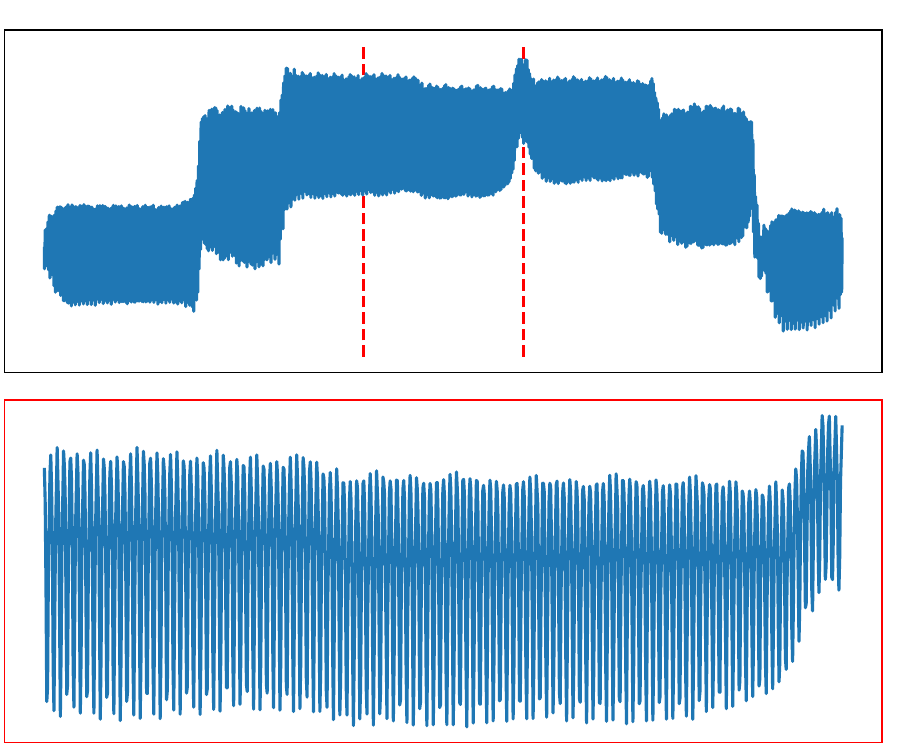} &
        \includegraphics[width=.225\textwidth]{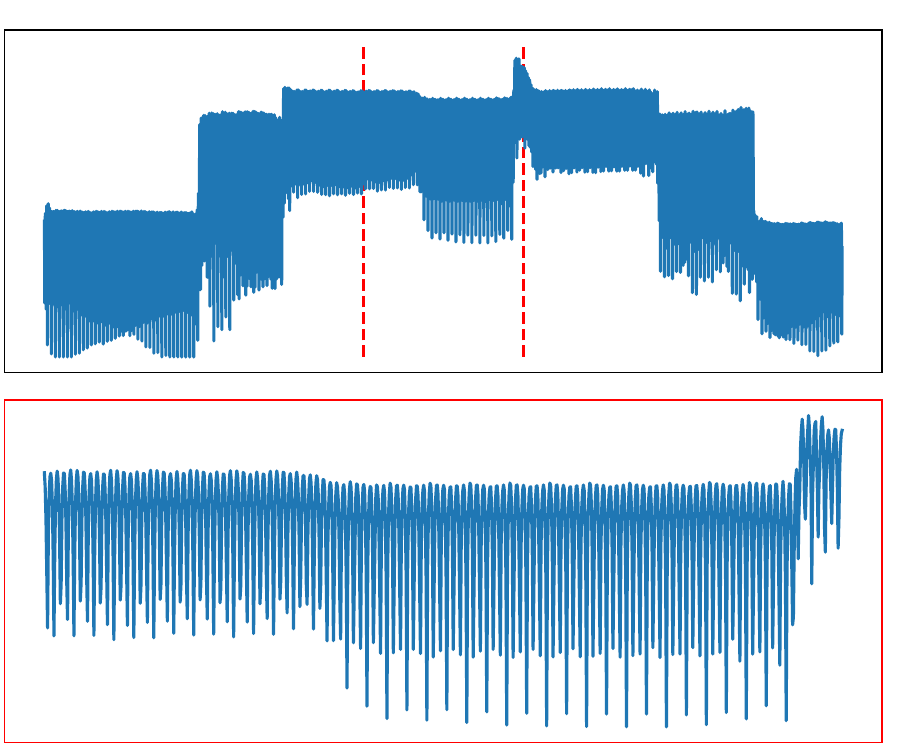} &
        \includegraphics[width=.225\textwidth]{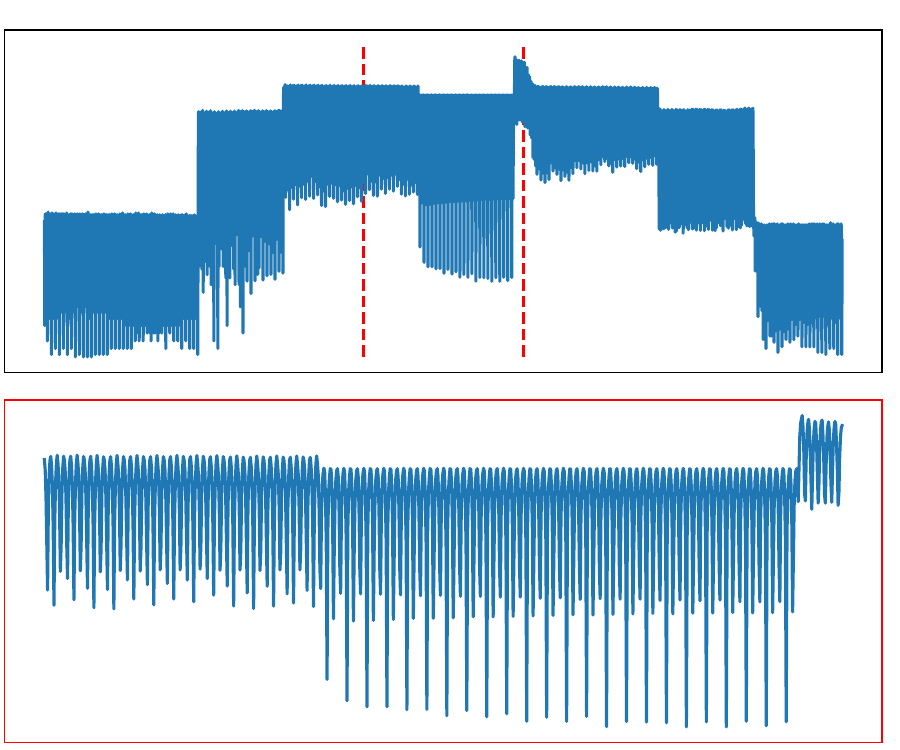}\\
        \includegraphics[width=.225\textwidth]{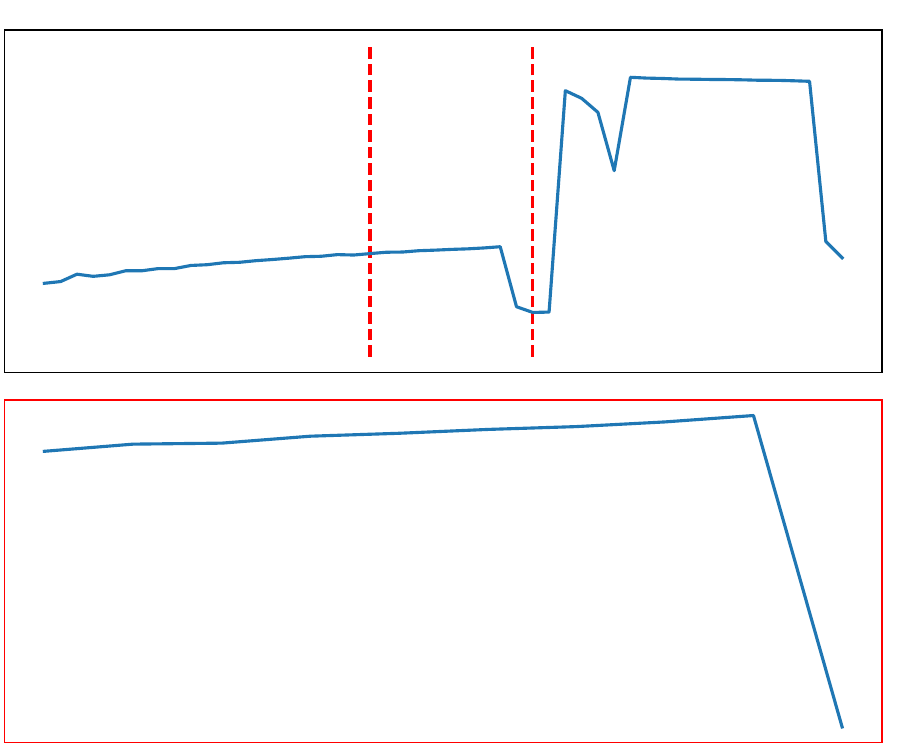} &
        \includegraphics[width=.225\textwidth]{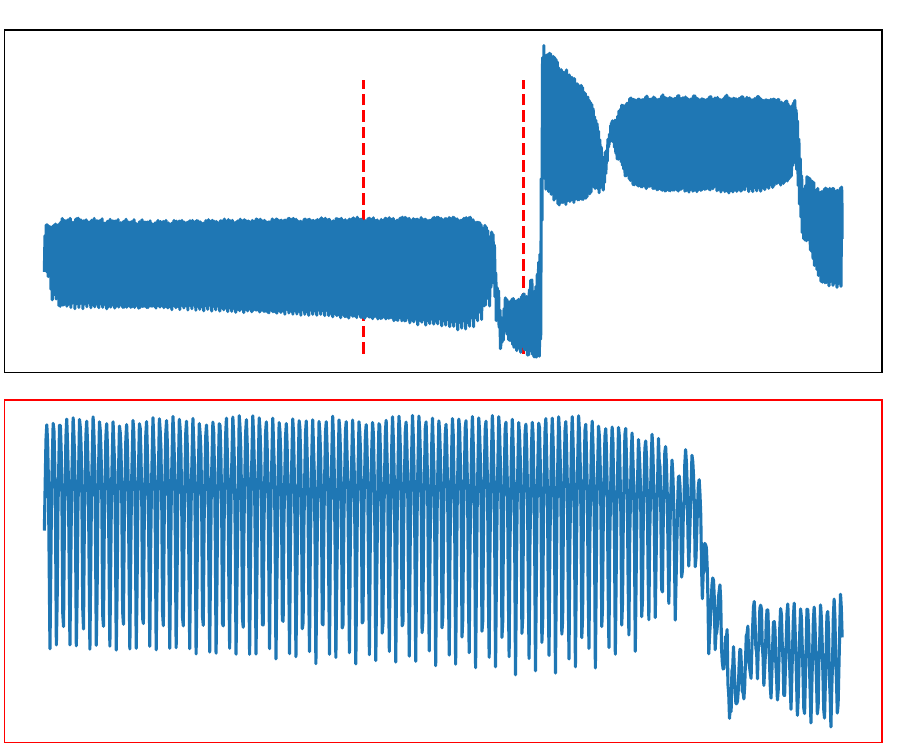} &
        \includegraphics[width=.225\textwidth]{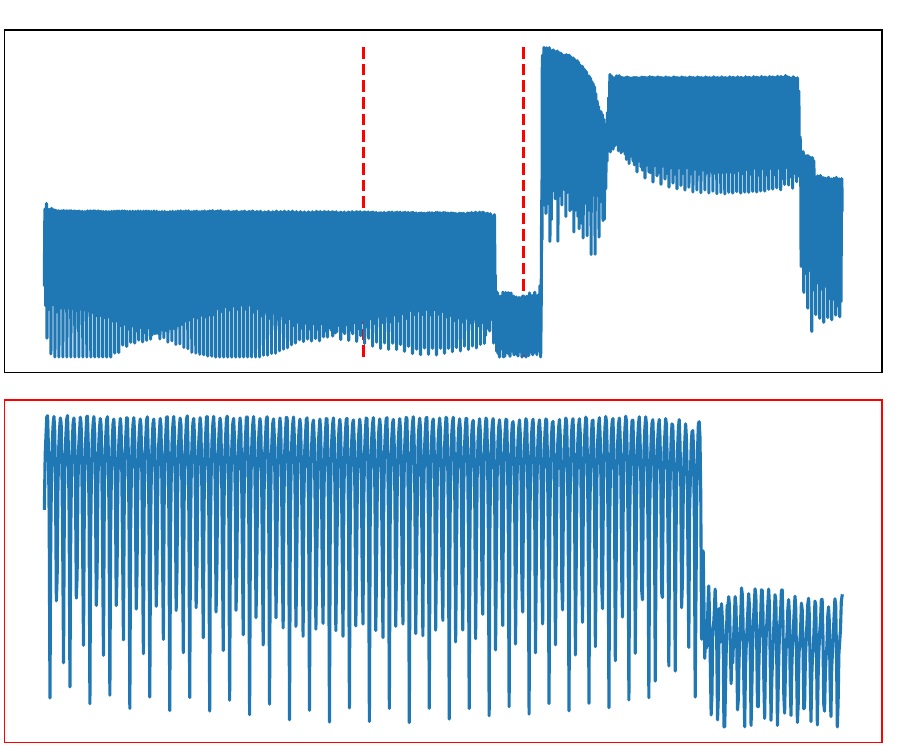} &
        \includegraphics[width=.225\textwidth]{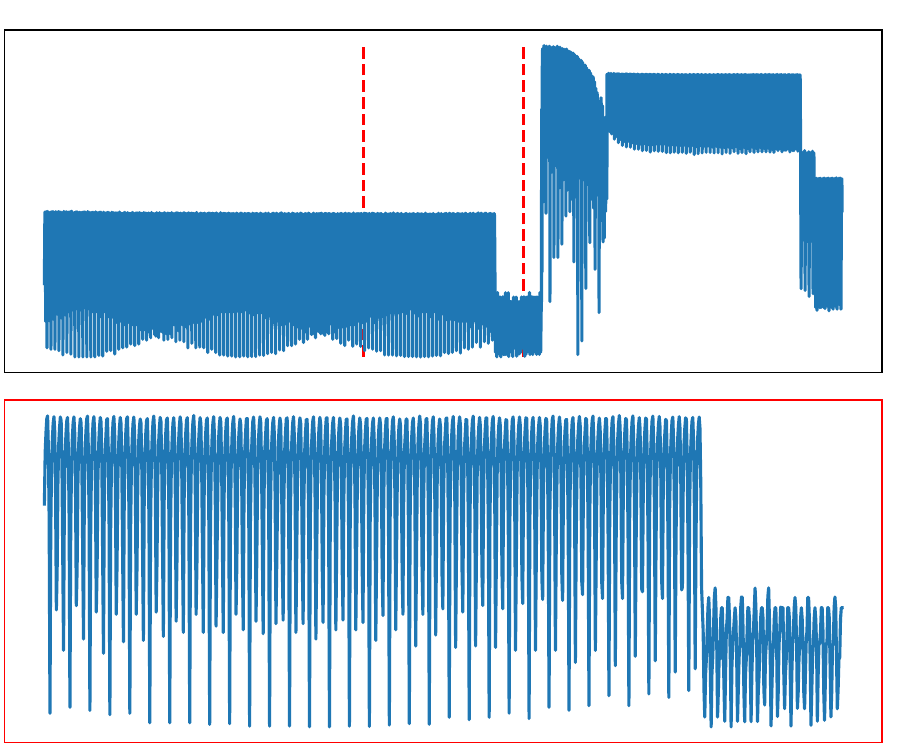}\\
        Linear Interpolation &  AudioNet   & SRPNet (ours)  & Ground Truth
    \end{tabular}
    \vspace{-4mm}
    \caption{The SRP results of experiment with $f_l=10\mathbf{Hz}$ and $\alpha=100$.}
    \label{fig:10-1000}
    \vspace{-4mm}
\end{figure*}
In the proposed method, the objective function consists of three parts.
We first employ the $l_2$-norm loss:
$$\mathcal{L}_2(y, y')=\|y-y'\|^2_2,$$
where $y'$ indicates the generated high-resolution data.
However, because the $l_2$-norm loss contains square operation, the gap between the larger error and the smaller error will be enlarged.
In other words, the $l_2$-norm will punish the larger error strongly and will tolerate the smaller error.
This brings difficult to reconstruct the waveform of the smart grid signal.
What is more important is that we want the super-resolved data could help the subsequent downstream applications such as recognition and monitoring.
In this case, the reconstruction of waveforms is practically essential.

In order to better reconstruct the waveforms and make the super-resolved data suitable for the subsequent monitoring application, we adapt a CNN based recognition network after the generation network as a recognition-based loss to explicitly constrain the reconstruction of waveforms.
By calculating losses through the recognition network, it will force the SRPNet to generate waveforms that can be recognized by the recognition network correctly.
This recognition loss is formulated as the cross-entropy loss between the predicted label and the ground truth label:
$$\mathcal{L}_{Rec}(y, y')=-\sum R(y')\log a+(1-R(y'))\log (1-a),$$
where $R$ is the CNN recognition network, and $a$ is the label vector indicating the running electric appliance of $x$.
For each appliance, the label $a$ is either 1 or 0, where the former indicates this appliance is sunning, and the later indicates the opposing situation.
$R$ is pretrained with the ground truth high-resolution data and is able to capture the crucial features that are important for recognition.
We adapt cross-entropy loss to train the $R$ network.

In addition to the above losses, we also adopt adversarial loss \cite{srgan} to learn the HR data manifold for the long sequence from the global perspective.
The adversarial loss makes the generated long sequences close to the target manifold.
The adversarial training strategy contains the generator and a discriminator.
%
%
The goal of the discriminator is to try to separate the signals generated by the generator from the real ones.
Ideally, the generator can generate signals with a realistic waveform that can deceive a well-trained discriminator.
Inspired by ESRGAN \cite{ESRGAN}, we change the standard GAN loss to Relativistic GAN loss \cite{jolicoeur2018relativistic}.
Different from the standard discriminator $D$, which estimates the probability that one input $y$ is on the target manifold, a relativistic discriminator tries to predict the probability that a real data $y$ is relatively more realistic than a fake one $y'$.
The standard discriminator can be expressed as $D(y) = \mathnormal{sigmoid}(C(y))$, where $C(y)$ is the discriminator output before any transformation.
Then, the used Relativistic Discriminator $D_{Ra}$ is formulated as:
$$
D_{Ra}(y, y') = \mathnormal{sigmoid}(C(y)-\mathbb{E}_{y'} [C(y')]),
$$
where $\mathbb{E}_{y'}$ represents the operation of taking the average for all fake HR data in one mini-batch.
And the discriminator loss is then formulated as:
$$
\mathcal{L}_{D}^{Ra}=-\mathbb{E}_{y}[\log(D_{Ra}(y,y'))]-\mathbb{E}_{y'}[\log(1-D_{Ra}(y,y'))].
$$
The adversarial loss for the SRP network $G(\cdot)$ is in a symmetrical form:
$$
\mathcal{L}_{G}^{Ra}=-\mathbb{E}_{y}[\log(1-D_{Ra}(y,y'))]-\mathbb{E}_{y'}[\log(D_{Ra}(y,y'))].
$$
In this adversarial loss, both $y$ and $y'$ are contained explicitly. Thus the Relativistic GAN loss can consistently produce gradients close to the data manifold without being affected by the over-training of the discriminator.

The overall loss function used for our full methods is:
$$
\mathcal{L}=\lambda_{sr}\cdot\mathcal{L}_2+\lambda_{adv}\cdot\mathcal{L}_{G}^{Ra}+\lambda_{rec}\cdot\mathcal{L}_{Rec},
$$
where $\lambda_{sr}$, $\lambda_{adv}$ and $\lambda_{rec}$ are the hyper-parameter of for content $l_2$-norm loss, the adversarial loss and recognition loss, respectively.

\begin{figure*}[!htb]
    \centering
    \begin{tabular}{cccc}
        \includegraphics[width=.225\textwidth]{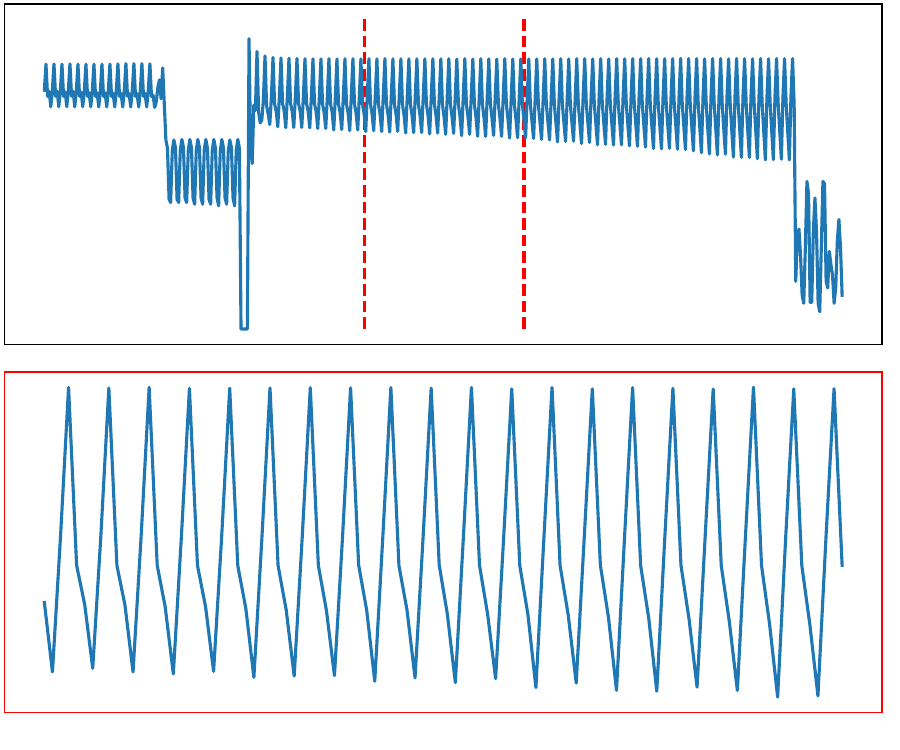} &
        \includegraphics[width=.225\textwidth]{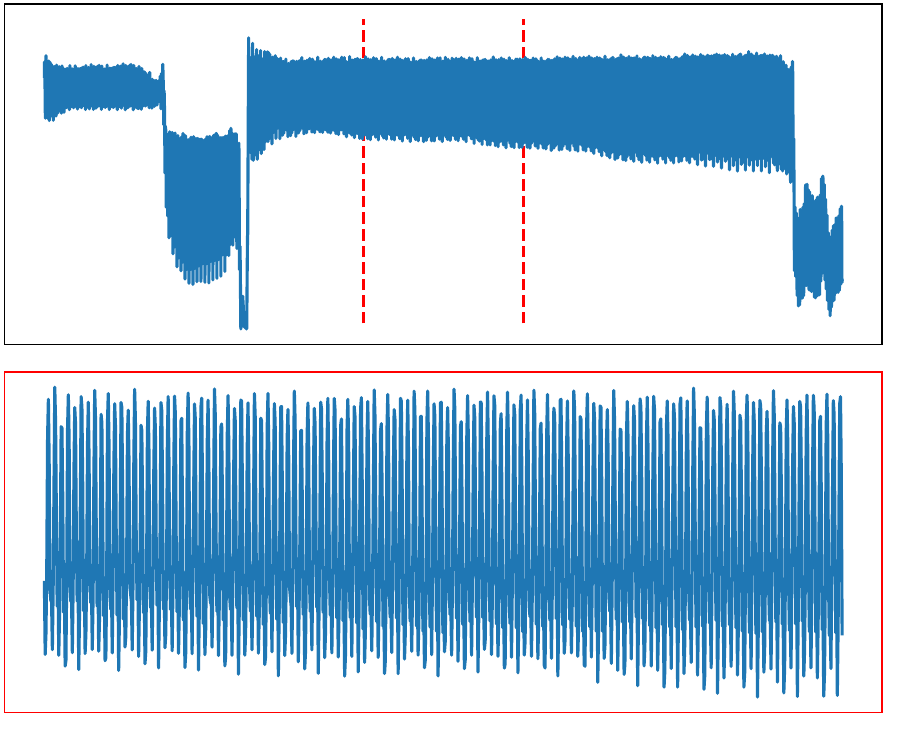} &
        \includegraphics[width=.225\textwidth]{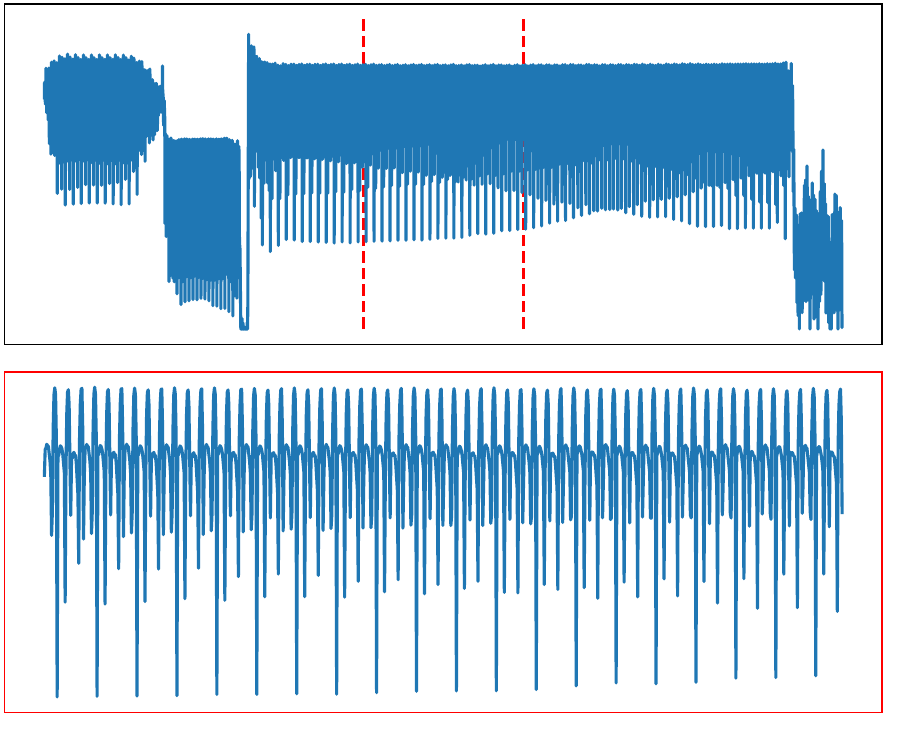} &
        \includegraphics[width=.225\textwidth]{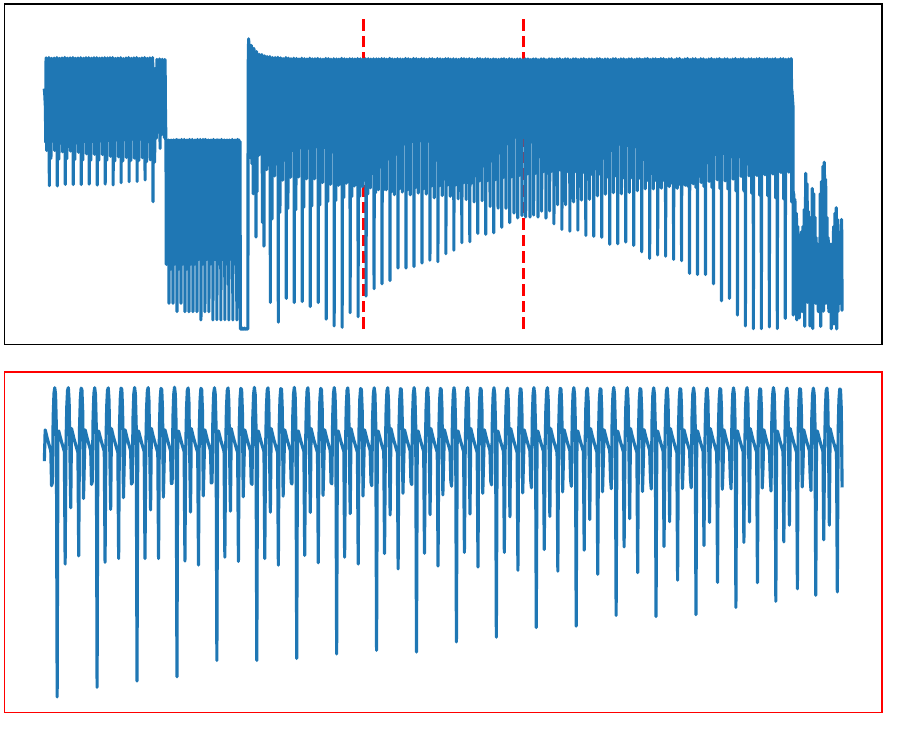} \\                    
        \includegraphics[width=.225\textwidth]{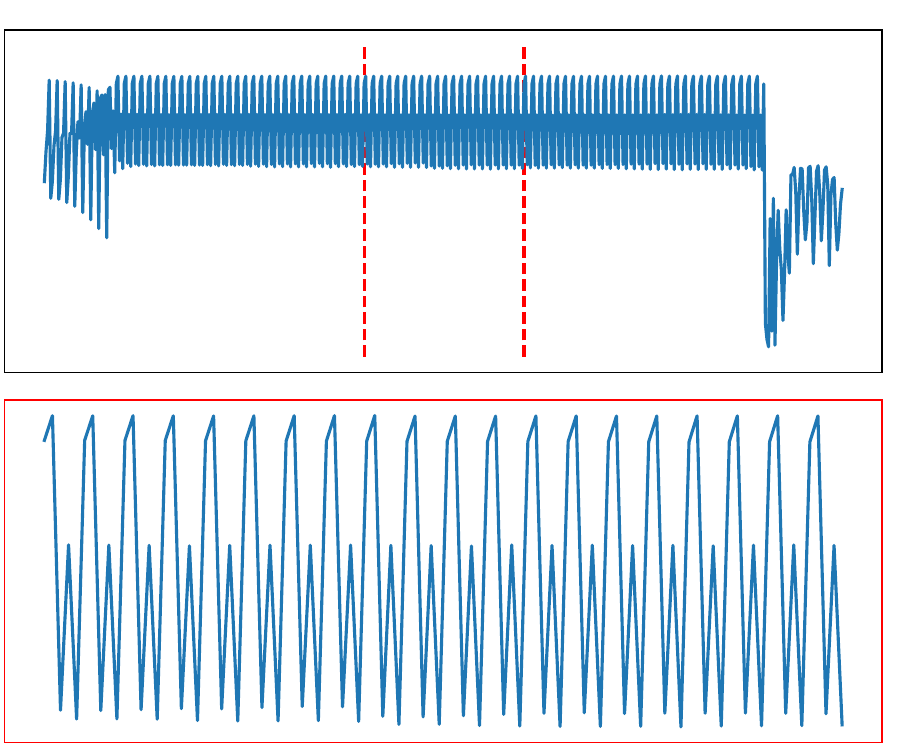} &
        \includegraphics[width=.225\textwidth]{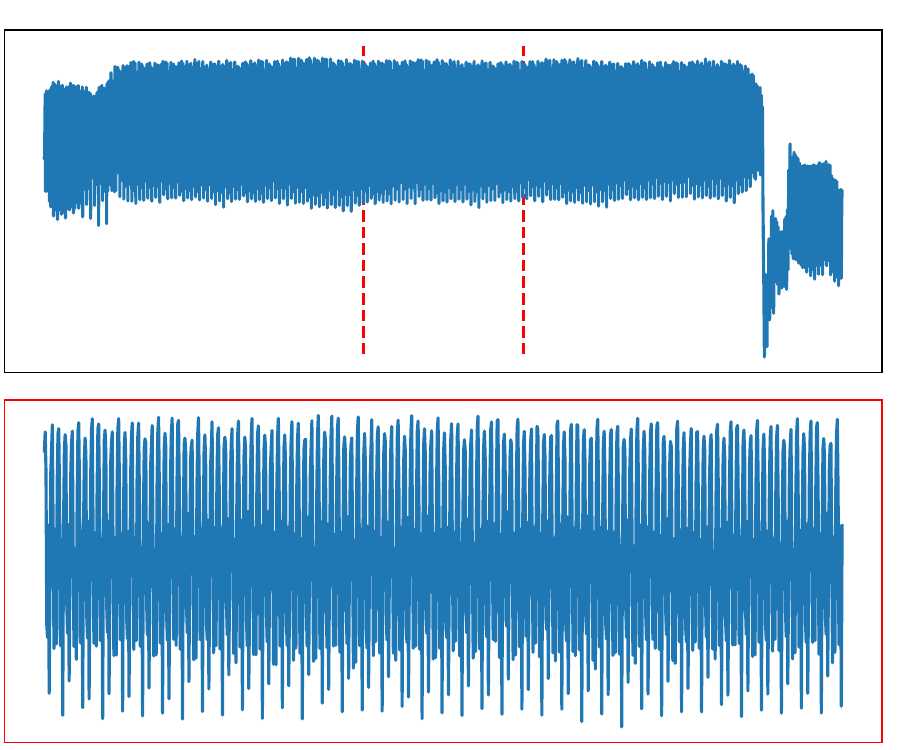} &
        \includegraphics[width=.225\textwidth]{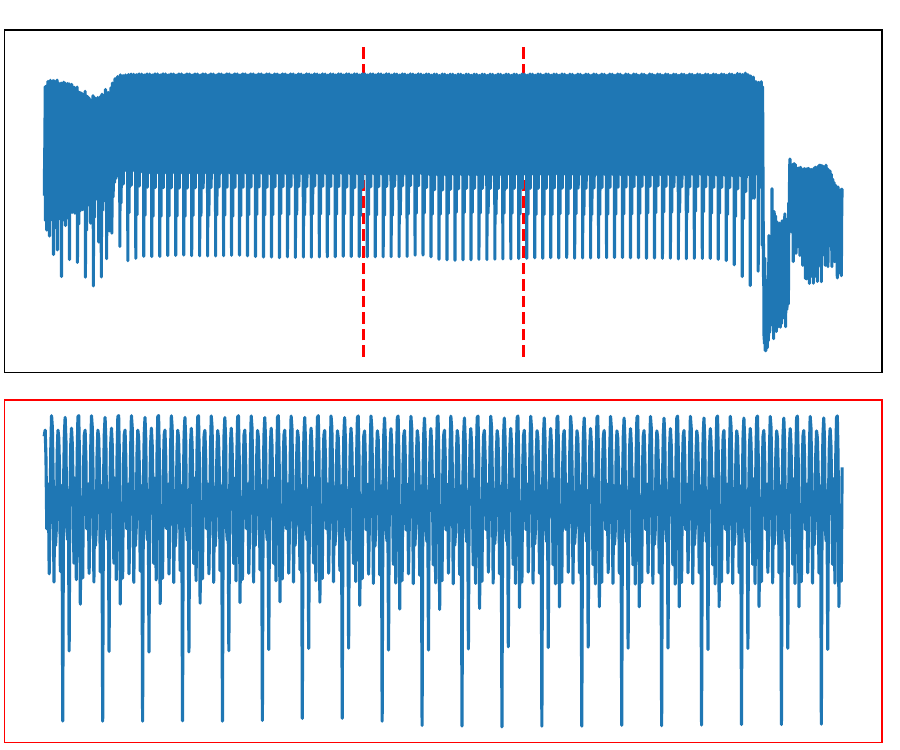} &
        \includegraphics[width=.225\textwidth]{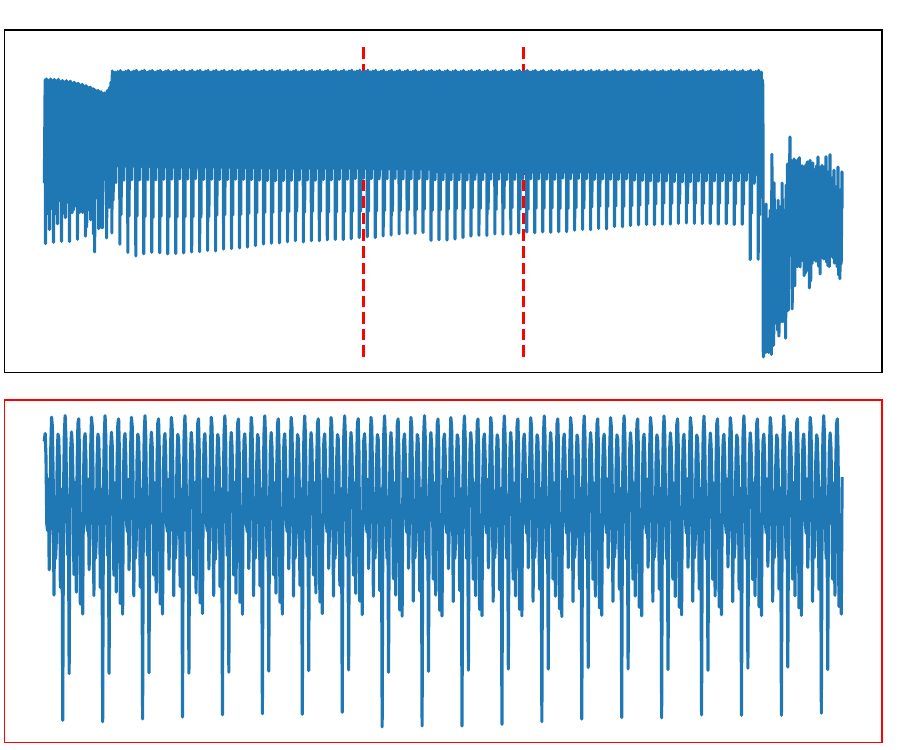} \\

        Linear Interpolation & AudioNet   & SRPNet (ours)  & Ground Truth
    \end{tabular}
    \vspace{-4mm}
    \caption{The SRP results of experiment with $f_l=100\mathbf{Hz}$ and $\alpha=10$.}
    \label{fig:100-1000}
    \vspace{-4mm}
\end{figure*}

\subsubsection{Network training.}
\label{sec:train}
For optimization, we use Adam \cite{kingma2014adam} with $\beta_1 = 0.9$ and $\beta_2 = 0.999$. The mini-batch size is set to $32$.
For each batch, the data are randomly cropped into 10 seconds for the training.
The learning rate is initialized as $1 \times 10^{-4}$ for the first $1 \times 10^6$ mini-batch updates and then fine tuned with learning rate of $1\times 10^{-6}$ for another $1 \times 10^6$ mini-batch updates.
For the hyper-parameter of objective function, we use $\lambda_{sr} = 1$, $\lambda_{adv} = 0.01$ and $\lambda_{rec} = 0.01$ for the best performance.
The settings of the loss hyper-parameter will be discussed in the ablation study section.
We implement our models with the PyTorch framework and train them using NVIDIA Titan Xp GPUs.
The entire training process took about twenty hours.

\subsection{Dataset and Data Preparation}
We use the simulated smart meter data to train the proposed SRPNet and test it on real-world data.
For the simulation data, we simulated the working state of electrical appliances in houses and then combined the corresponding electrical load data into a final smart meter data for a house according to the working state.
The used electrical load data of appliance in different states is provided in the Plug-Level Appliance Identification Dataset (PLAID) \cite{de2017handling}.
PLAID can be used in appliance identification; it provides $1793$ high resolution ($30000\mathbf{HZ}$) meter data instances of $82$ different appliances which belong to $11$ different appliance types.
In the experiment, we prepare $14000$ high-frequency data samples of length $30\mathbf{s}$ for training, and another $2000$ data were used for validation.
For the testing data, we collect real-world electricity load data where the main appliances are similar to the appliances in training data.
The testing set contains 130 hours of data.

Since the electrical load data has a very large dynamic range ($10^{-3}$ - $10^{6}$), it is difficult for the neural network to directly process the input and output with such a large dynamic range.
We preprocess the original large dynamic meter data with the following formula 
$$\tilde{x}=\log_{100}(x\times10^3+1).$$
This formula is monotonous and ensured $\tilde{x}$ to be positive.
It well preserves the fluctuations in data.
The high and low-frequency data used for training is based on the degradation model described in the problem formulation section.
Since the meter records the instantaneous current and voltage values, we assume the sampled data given by the high-frequency meter and low-frequency meter are approximately equal.
Thus, we employ the Nearest Neighbor (NN) down-sampling as the degradation function.
Then, Gaussian noise with a $\sigma=0.01$ was added to the down-sampled low-frequency data.

\subsection{SRP Results}
In order to study the characteristics and effectiveness of the proposed SRP method, we conducted experiments under three different settings:
(1) the low-res data with frequency $f_l=10\mathbf{Hz}$ and an SRP factor of $\alpha=10$;
(2) $f_l=100\mathbf{Hz}$ and $\alpha=10$;
(3) $f_l=10\mathbf{Hz}$ and of $\alpha=100$.
    
\subsubsection{Quantitative Metrics}\label{sec:metric}
To quantify the comparison, we use two metrics:
Signal to noise ratio (SNR) and The Log-spectral distance (LSD) \cite{LSD}.
SNR is a commonly used metric in the signal processing literature.
Given a reference signal $y$ and an approximation $y'$, the SNR is defined as
$$SNR(y,y')=10\times\log\frac{\|y\|^2_2}{\|y-y'\|^2_2}.$$
The LSD measures the reconstruction quality of individual frequencies as follows
$$LSD(y,y')=\frac{1}{L}\sum_{l=1}^L\sqrt{\frac{1}{K}\sum_{k=1}^K\big(X(l,k)-X'(l,k)\big)^2},$$
where $X$ and $X'$ are the log-spectral power magnitudes of $y$ and $y'$, respectively.
These are defined as $X=\log\|S\|^2$, where $S$ is the short-time Fourier transform of the signal.
We use $l$ and $k$ index frames and frequencies, respectively.
By combining the above metrics, we can evaluate the performance of temporal SRP methods.
    
\begin{table*}[!htb]
\begin{center}
    \newcommand{\tabincell}[2]{\begin{tabular}{@{}#1@{}}#2\end{tabular}}
    \begin{tabular}{p{2.8cm}ccccccccc}
        \hline
        \multirow{2}{*}{Method} & \multicolumn{2}{c}{$f_l=10\mathbf{Hz}$, $\alpha=10$} & \multicolumn{2}{c}{$f_l=100\mathbf{Hz}$, $\alpha=10$} & \multicolumn{2}{c}{$f_l=10\mathbf{Hz}$, $\alpha=100$}\\
        & SNR $\uparrow$ &  LSD $\downarrow$  & SNR $\uparrow$    &  LSD $\downarrow$  & SNR $\uparrow$ & LSD $\downarrow$ \\
        \hline
        \hline
        Linear Interpolation &
        $17.86$ & $8.136$ & $17.93$  &  $5.685$ & $17.43$ &  $13.03$\\
        Li \textit{et al.} \cite{Li2015A} &
        $20.21$ & $2.718$ & $28.24$  &  $2.021$ & $20.08$ &  $2.569$\\            
        AudioNet &
        $25.93$ & $2.253$ & $35.83$ & $1.279$ & $23.21$ & $3.109$\\
        SRPNet ($l_2$) &
        $26.05$ & $1.898$ & $33.41$ & $1.460$ & $23.30$ & $2.371$  \\
        SRPNet (adv) &
        $26.54$ & $1.852$ & $35.27$ & $1.264$ & $24.33 $ & $2.218$ \\
        SRPNet (full model) &
        $\mathbf{27.90}$ & $\mathbf{1.673}$ & $\mathbf{36.18}$ & $\mathbf{1.221}$ & $\mathbf{24.39}$ & $\mathbf{2.200}$  \\    
        \hline
        \hline
    \end{tabular}

\end{center}
    \caption{The quantitative comparison of linear interpolation, AudioNet \cite{kuleshov2017audio} and SRPNet (SNR / LSD). SRPNet ($l_2$) means using only $l_2$-norm loss, SRPNet (adv) means training SRPNet with adversarial loss, and SRPNet (full model) indicates our full model including recognition-based loss. $\uparrow$ means the higher the better while  $\downarrow$ means the lower the better.}
    \vspace{-6mm}
    \label{tab:srp_results}

\end{table*}%

\subsubsection{Quantitative Comparison}\label{sec:quantitative}
We compared our method with linear interpolation, Li \textit{et al.} \cite{Li2015A}, and AudioNet \cite{kuleshov2017audio}.
We also compared the performance of SRPNet using different losses.
The quantitative comparison results are shown in the \tablename~\ref{tab:srp_results}.
It is obvious that the SRP method is significantly better than the existing methods.
This experiment also demonstrates the effectiveness of the proposed recognition-based loss and adversarial loss.
We can also find that the experiment of input frequency with $100\mathbf{hz}$ and an SRP factor of $10$ has the best SRP result.
Because the high-frequency data contains more information on describing the real system state.
Thus, it is easy for the generative model to infer the data before degradation.
When the frequency of the input data becomes lower, the real system state is more difficult to estimate, and it is more difficult to recover high-frequency information therefrom.
We can also see that when the SRP factor becomes $100$, although the SRP can still recover the high-frequency data to a certain extent, it suffered more severe distortion than the experiment with an SRP factor of $10$.
These experiments demonstrate that: (1) when the input data has a lower frequency, it is more difficult to perform SRP; (2) it is more difficult to perform SRP with a larger super-resolution factor.

\subsubsection{Qualitative Comparison}\label{sec:qualitative}
Since the existing traditional (not deep learning-based) time series upsampling methods (such as compressed sensing and dictionary learning) cannot perform SRP with SR factors of 10 or even 100, the comparison is conducted with linear interpolation and AudioNet \cite{kuleshov2017audio}. 
Some SRP test results are shown in \figurename~\ref{fig:100-1000} and \figurename~\ref{fig:10-1000}.
For each sample, the first line of each case is the overview of a $5$ second time slice, and the second line is the zoom-in view of $1$ second marked by red lines in the overview.
As can be seen, the interpolation method can not recover the missing details in the high-frequency data due to the loss of information during the degradation process.
The AudioNet try to reconstruct waveforms with some high-frequency signals, yet the reconstruction is unreliable.
Compared with our method, AudioNet can only restore the rough shape of the waveform; the details of the waveform are difficult to recover. 
However, the proposed SRP method can restore the waveforms and details effectively.

\begin{figure}[t]
    \centering
    \begin{tabular}{ccc}
        \includegraphics[width=.47\textwidth,height=.07\textwidth]{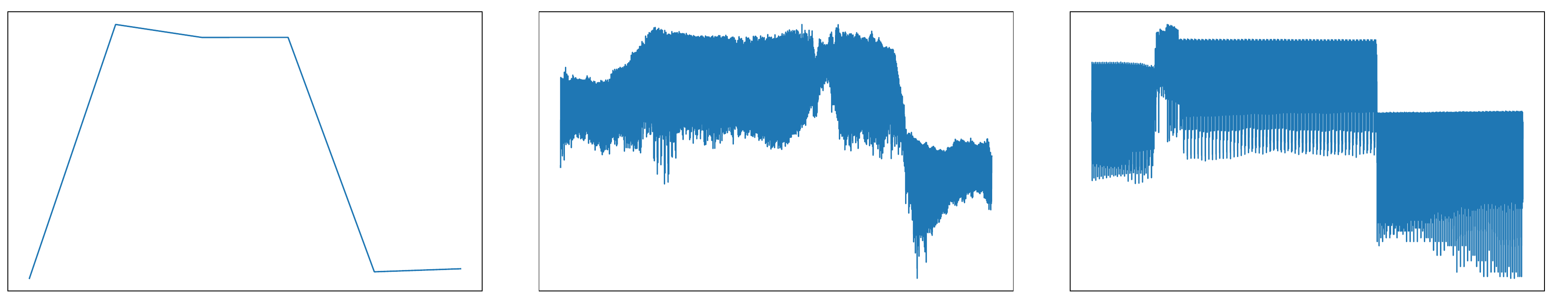} \\
        \includegraphics[width=.47\textwidth,height=.07\textwidth]{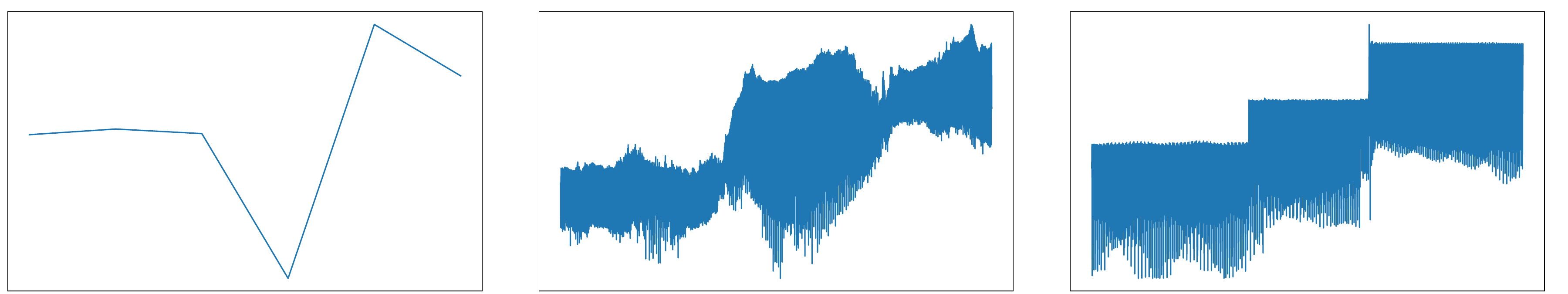} &\\        \includegraphics[width=.47\textwidth,height=.07\textwidth]{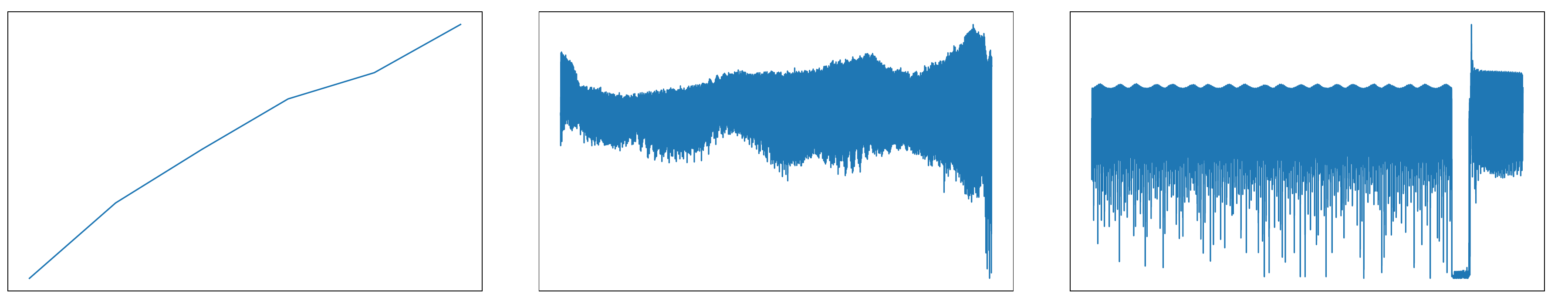} &\\
        \includegraphics[width=.47\textwidth,height=.07\textwidth]{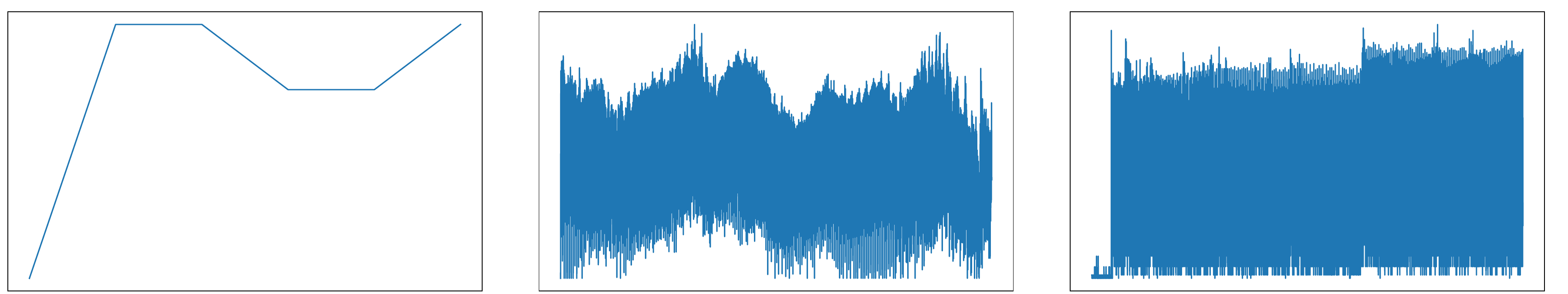} &\\        \includegraphics[width=.47\textwidth,height=.07\textwidth]{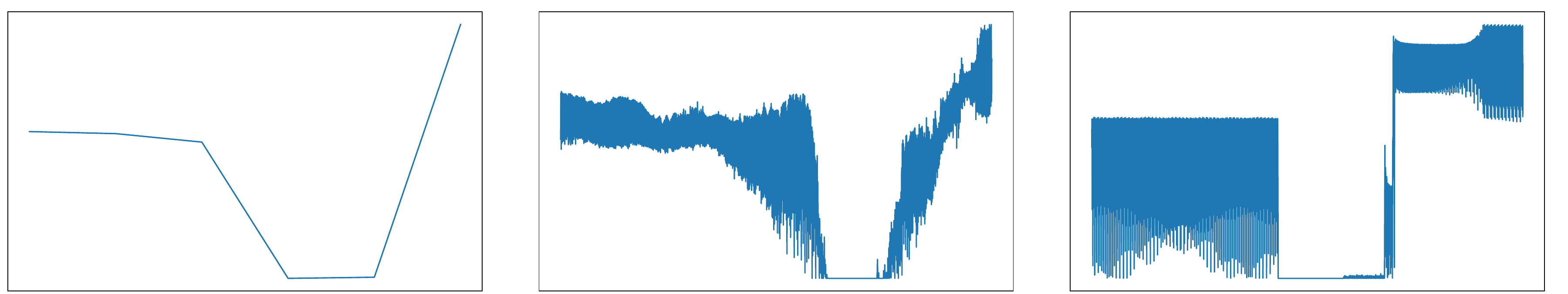} &\\
        
    \end{tabular}
    
    \begin{tabular}{p{2.2cm}<{\centering}p{2.67cm}<{\centering}p{2.05cm}<{\centering}}
        Linear Interpolation & SRPNet (ours) & Ground Truth
    \end{tabular}
    \caption{The SRP results of experiment with $f_l=1\mathbf{Hz}$ and $\alpha=1000$.}
    \label{fig:hard case}
    \vspace{-4mm}
\end{figure}

Then we focus on some hard cases for the SRP problem: the low-resolution data with frequency $f_l=1\mathbf{Hz}$ and an SR factor of $\alpha=1000$.
Some examples are shown in \figurename~\ref{fig:hard case}.
The low-frequency data are 6 seconds slice of $1\mathbf{Hz}$ smart meter data, which means there are only 6 data points.
It is very difficult to restore the original waveform with extremely limited information.
In this case, SRPNet can only recover part of the waveform information.
The reconstructed signals have greater distortion compare to the ground truth.
However, this experiment still shows the potential of applying SRP with a large SRP factor.

\subsubsection{Ablation Study.}\label{sec:ablation}
\begin{figure*}[htb]
    \centering
    \begin{tabular}{ccccc}
        \includegraphics[width=.18\textwidth]{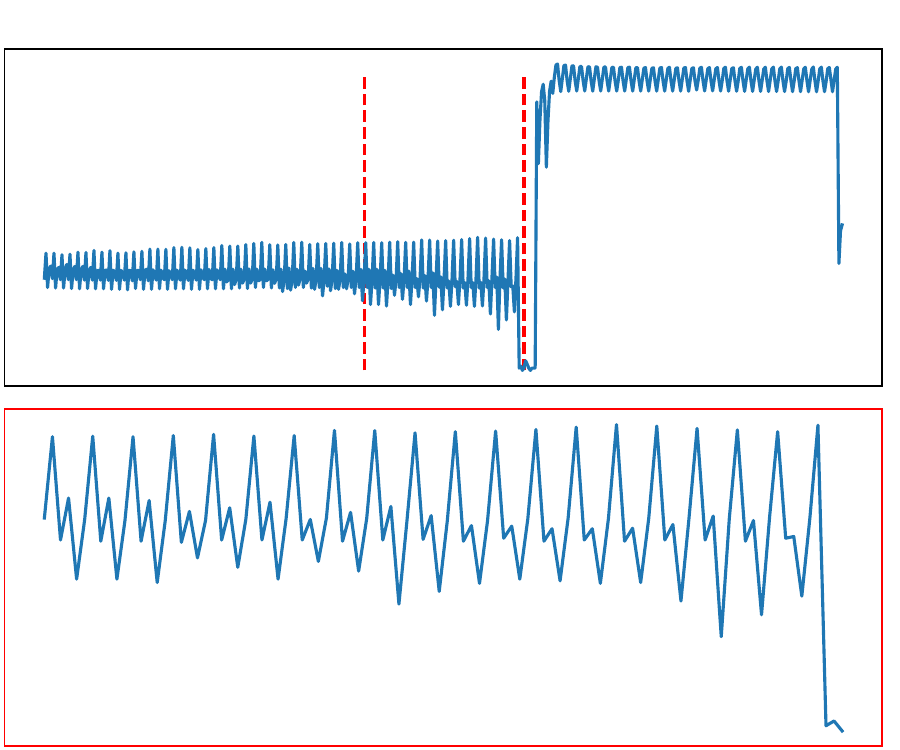} &
        \includegraphics[width=.18\textwidth]{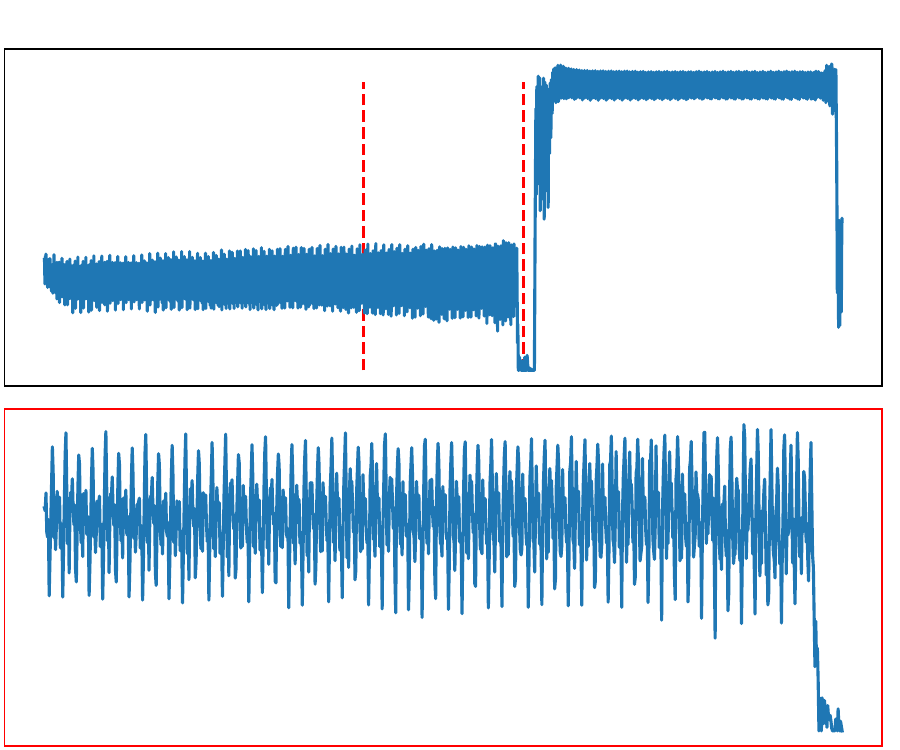} &
        \includegraphics[width=.18\textwidth]{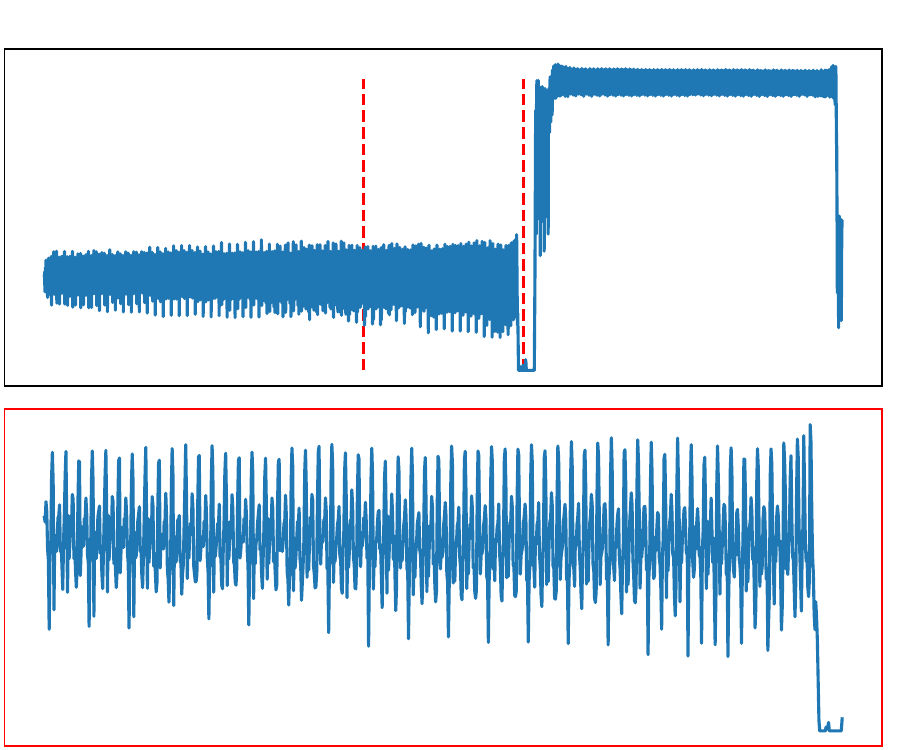} &
        \includegraphics[width=.18\textwidth]{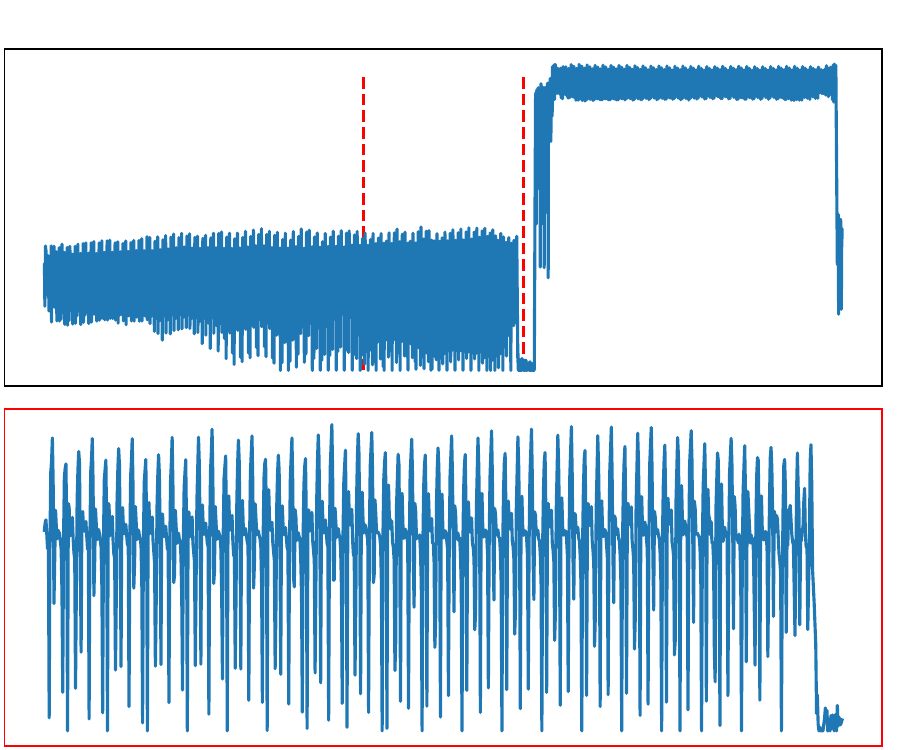} &
        \includegraphics[width=.18\textwidth]{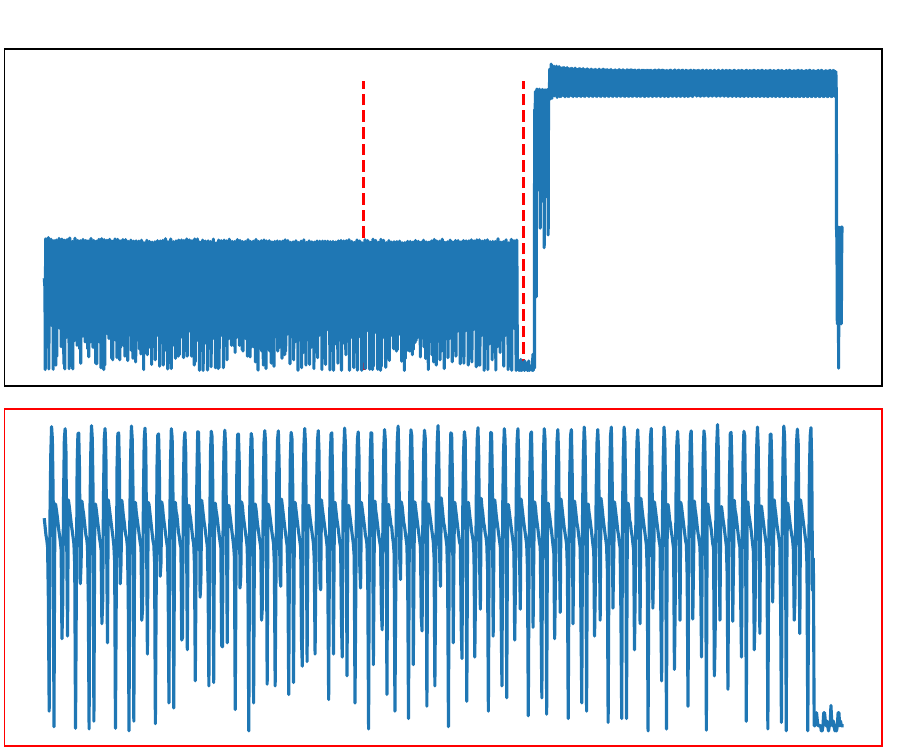}\\
        \includegraphics[width=.18\textwidth]{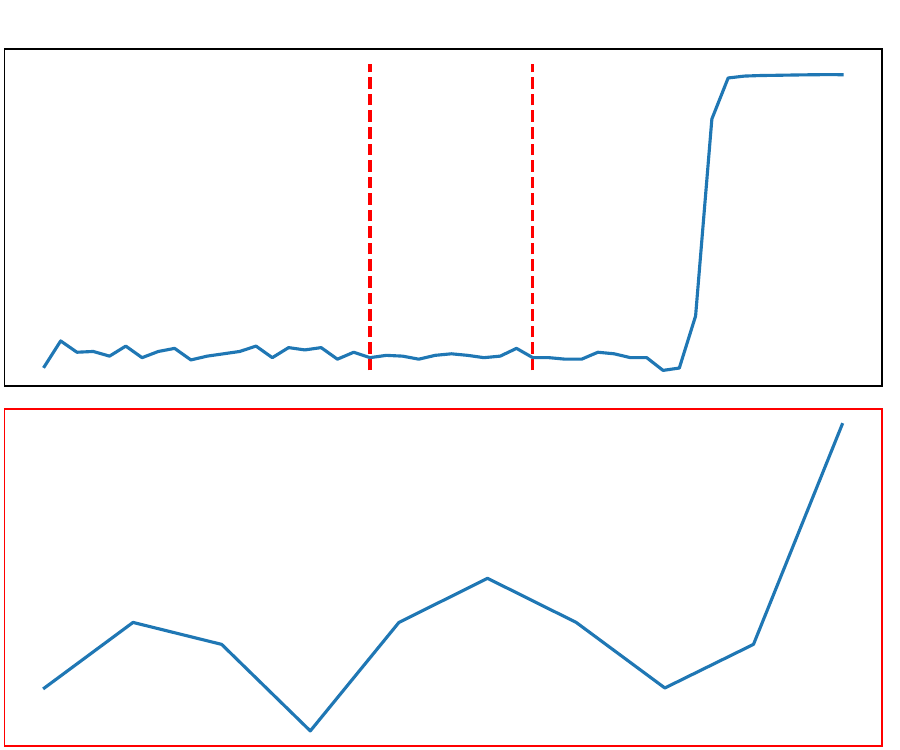} &
        \includegraphics[width=.18\textwidth]{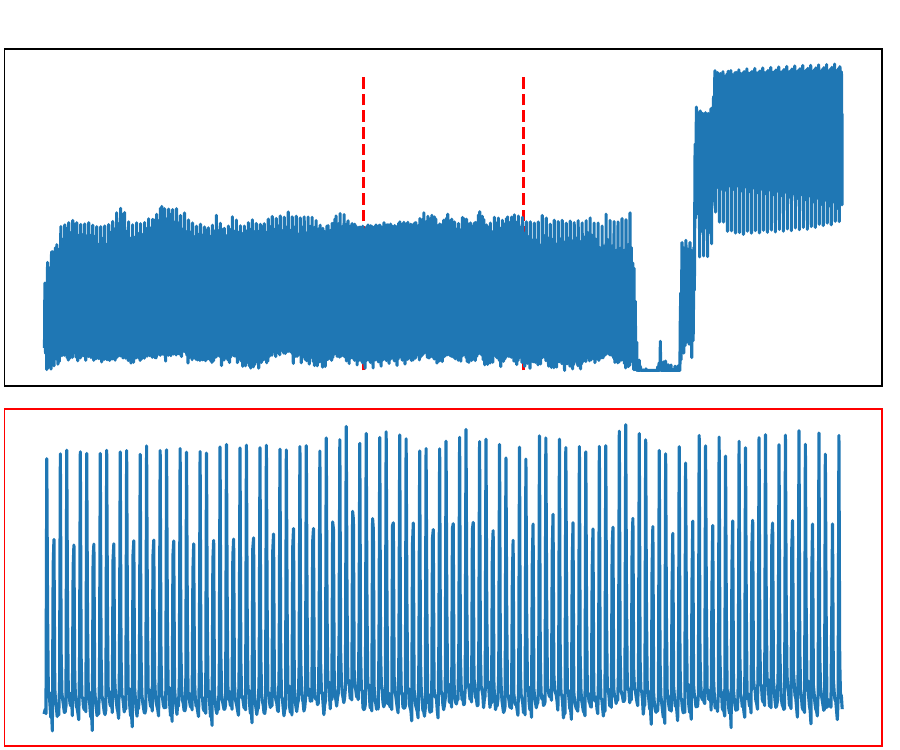} &
        \includegraphics[width=.18\textwidth]{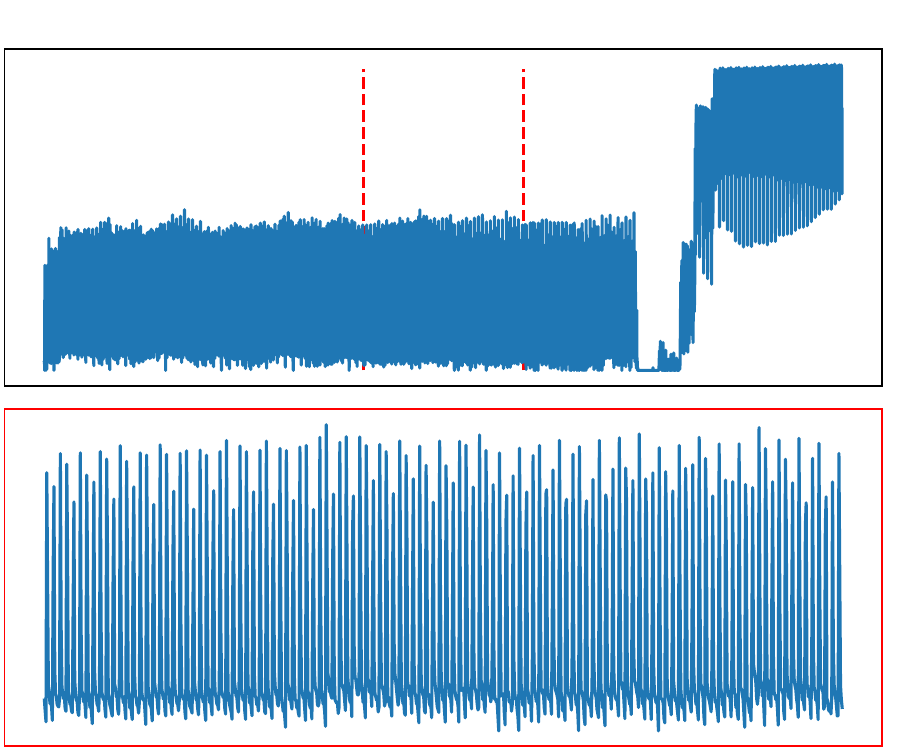} &
        \includegraphics[width=.18\textwidth]{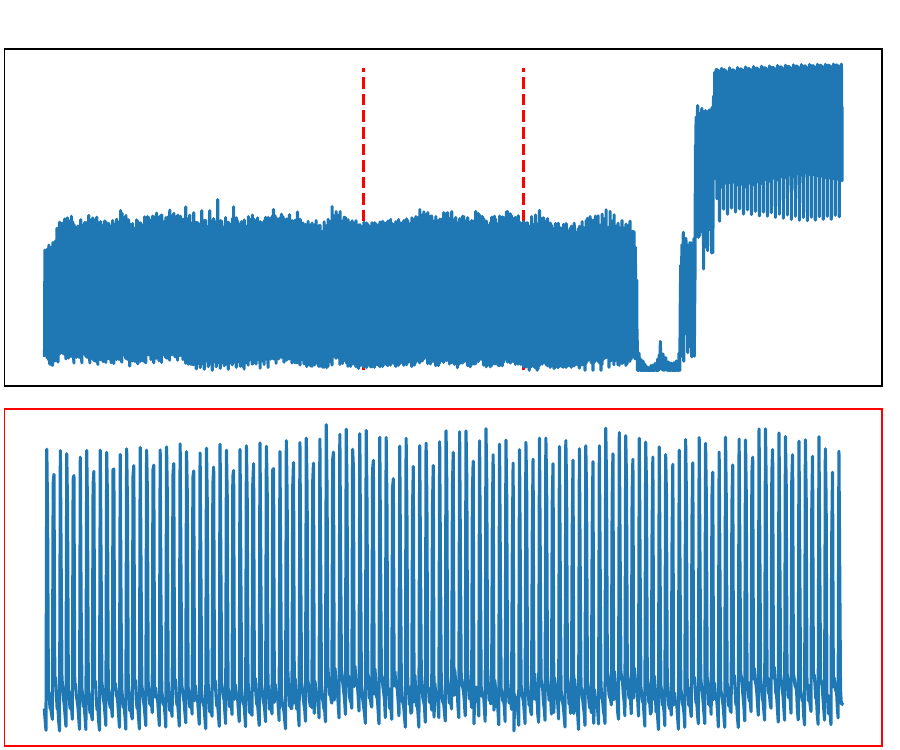} &
        \includegraphics[width=.18\textwidth]{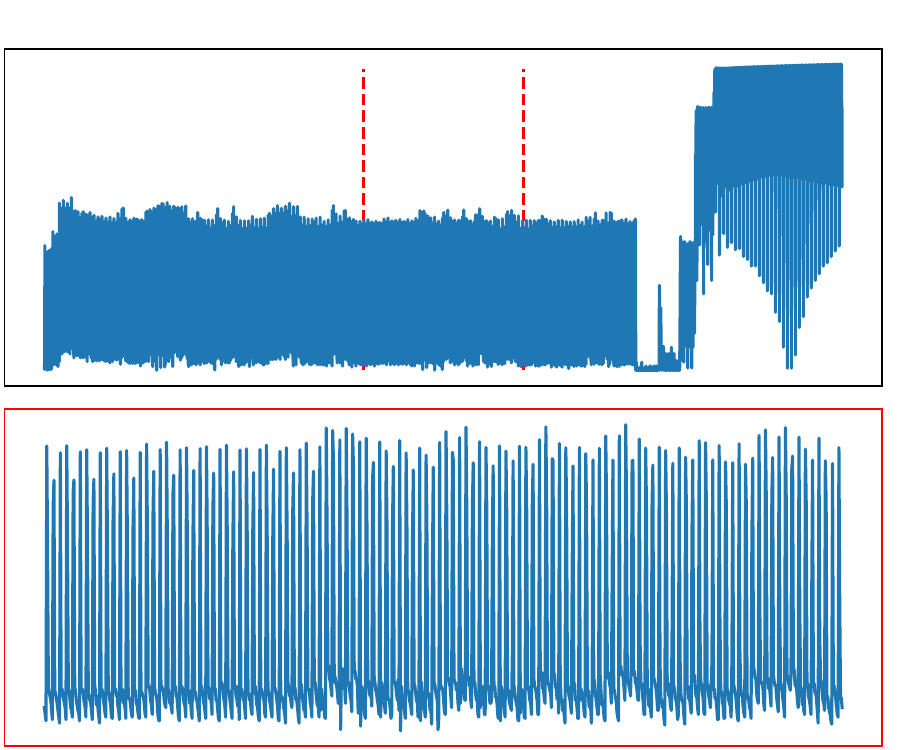}\\
        Linear Interpolation &  SRPNet ($l_2$) & SRPNet (adv) & SRPNet (full model)& Ground Truth
    \end{tabular}
    \caption{The first two rows are SRP results of experiment with $f_l=100\mathbf{Hz}$ and $\alpha=10$, the third and fourth rows are SRP results of experiment with $f_l=10\mathbf{Hz}$ and $\alpha=100$.}
    \label{fig:ablation}
\end{figure*}

To investigate the behaviors of these two losses, we conducted several ablation studies.
First, we trained the network with only $l_2$-norm loss as the baseline.
Then, we added adversarial loss and recognition-based loss with different hyper-parameters to show how they affect the results.
In this ablation study, the input frequency is $100\mathbf{hz}$, and the SRP factor is $10$.
The SRP results are shown in \figurename~\ref{fig:ablation}.
\tablename~\ref{tab:a1} shows the effect of adversarial loss.
We can observe that the results using adversarial loss show better performance, and when $\lambda_{adv}=0.01$, it achieves the best performance.
\tablename~\ref{tab:hyperparameters} shows the effect of using recognition-based loss.
As one can see, the performance of the solution with recognition loss is generally better than the solution without recognition loss, and the performance is not sensitive to hyper-parameters.
We can further observe that by combining adversarial loss and recognition loss, we can achieve the best performance when $\lambda_{adv}=0.01$ and $\lambda_{rec}=0.01$.

\begin{table}[t]
\begin{center}
    \newcommand{\tabincell}[2]{\begin{tabular}{@{}#1@{}}#2\end{tabular}}
    \begin{tabular}{p{2.9cm}ccccccccc}
        \hline
        \multirow{2}{*}{Method} & \multicolumn{2}{c}{$f_l=10\mathbf{Hz}$, $\alpha=10$} \\
        & SNR $\uparrow$ &  LSD $\downarrow$  \\
        \hline
        \hline
        $\lambda_{adv} = 0 $  & $26.05$ & $1.898$ \\
        \hline
        $\lambda_{adv} = 0.1 $ &
        $ 26.28 $  &  $1.889 $ \\
        $\lambda_{adv} = 0.01 $ &
        $ \mathbf{26.54} $  &  $ \mathbf{1.852} $  \\
        $\lambda_{adv} = 0.001 $ &
        $ 26.48 $  &  $ 1.886 $ \\
        $\lambda_{adv} = 0.0001 $ &
        $ 26.23 $ &  $ 1.893 $ \\
        \hline
        \hline
    \end{tabular}
\end{center}
    \caption{The quantitative comparison of different hyper parameters setting of SRPNet, $\lambda_{sr}$ defaults to 1. $\uparrow$ means the higher the better while  $\downarrow$ means the lower the better.}
    \label{tab:a1}
    \vspace{-8mm}
\end{table}    

\begin{table}[t]
\begin{center}
    \newcommand{\tabincell}[2]{\begin{tabular}{@{}#1@{}}#2\end{tabular}}
    \begin{tabular}{p{4.3cm}ccccccccc}
        \hline
        \multirow{2}{*}{Method} & \multicolumn{2}{c}{$f_l=10\mathbf{Hz}$, $\alpha=10$} \\
        & SNR $\uparrow$ & LSD $\downarrow$  \\
        \hline
        \hline
        $\lambda_{adv} = 0,\lambda_{rec} = 0 $  & $26.05$ & $1.898$ \\
        \hline            
        $\lambda_{adv} = 0.1, \lambda_{rec} = 0.1 $ &
        $27.32$  & $1.785$ \\
        $\lambda_{adv} = 0.1, \lambda_{rec} = 0.01 $ &
        $27.72$  & $1.730$  \\
        $\lambda_{adv} = 0.1,  \lambda_{rec} = 0.001 $ &
        $27.10$  & $1.783$ \\
        \hline
        $\lambda_{adv} = 0.01, \lambda_{rec} = 0.1 $ &
        $27.76$  & $1.695$ \\
        $\lambda_{adv} = 0.01, \lambda_{rec} = 0.01 $ &
        $\mathbf{27.90}$  & $\mathbf{1.673}$  \\
        $\lambda_{adv} = 0.01, \lambda_{rec} = 0.001 $ &
        $27.71$  & $1.722$ \\            
        \hline
        \hline
    \end{tabular}
\end{center}
\caption{The quantitative comparison of different hyper-parameters setting of SRPNet (full model), $\lambda_{sr}$ defaults to 1. $\uparrow$ means the higher the better while  $\downarrow$ means the lower the better.}
\label{tab:hyperparameters}
\vspace{-8mm}
\end{table}    

\begin{table*}[t]
    \centering
    \newcommand{\tabincell}[2]{\begin{tabular}{@{}#1@{}}#2\end{tabular}}
    \begin{tabular}{p{1.7cm}p{3.0cm}<{\centering}p{1.7cm}<{\centering}p{1.7cm}<{\centering}p{1.7cm}<{\centering}p{1.7cm}<{\centering}p{1.7cm}<{\centering}p{1.7cm}<{\centering}}
        \hline
        Method & Experiment &\tabincell{c}{trained on LF\\test on LF} & \tabincell{c}{trained on HF\\test on SRPNet \\ ($l_2$ only)} & \tabincell{c}{trained on HF\\test on SRPNet \\(full model)} & \tabincell{c}{trained on HF\\test on HF} & \tabincell{c}{Gain from\\SRPNet ($l_2$ only)}& \tabincell{c}{Gain from\\SRPNet \\ (full model)}\\
        \hline
        \hline
        \multirow{3}{*}{KNN}
        & $f_l=10\mathbf{Hz}$, $\alpha=10$ & $0.7703 $ & $0.7737 $ & $0.7779 $ & $0.7818 $ & $+0.0034 $ & $\mathbf{+0.0076 } $\\
        & $f_l=100\mathbf{Hz}$, $\alpha=10$ & $0.7818 $ & $0.7917 $ & $0.7926  $ & $0.7939 $ & $+0.0098 $ & $\mathbf{+0.0108 } $\\
        & $f_l=10\mathbf{Hz}$, $\alpha=100$ & $0.7703 $ & $0.7807 $ & $ 0.7898 $ & $0.7939 $ & $+0.0104 $ & $\mathbf{+0.0195 } $\\
        \hline
        \multirow{3}{*}{Decision Tree}
        & $f_l=10\mathbf{Hz}$, $\alpha=10$ & $0.6922 $ & $0.7056 $ & $ 0.7266  $ & $0.7332 $ & $+0.0134 $ & $\mathbf{+0.0344 } $\\
        & $f_l=100\mathbf{Hz}$, $\alpha=10$ & $0.7332 $ & $0.7401 $ & $ 0.7421 $ & $0.7502 $ & $+0.0069 $ & $\mathbf{+0.0089 } $\\
        & $f_l=10\mathbf{Hz}$, $\alpha=100$ & $0.6922 $ & $0.7225 $ & $ 0.7355 $ & $0.7502 $ & $+0.0303 $ & $\mathbf{+0.0433 } $\\
        \hline
        \multirow{3}{*}{MLP}
        & $f_l=10\mathbf{Hz}$, $\alpha=10$ & $0.8119 $ & $0.8202 $ & $0.8276 $ & $0.8332 $ & $+0.0083 $ & $\mathbf{+0.0157 } $\\
        & $f_l=100\mathbf{Hz}$, $\alpha=10$ & $0.8332 $ & $0.8340 $ & $ 0.8347 $ & $0.8390 $ & $+0.0008 $ & $\mathbf{+0.0015 } $\\
        & $f_l=10\mathbf{Hz}$, $\alpha=100$ & $0.8119 $ & $0.8198 $ & $0.8235 $ & $0.8390 $ & $+0.0079 $ & $\mathbf{+0.0116 } $\\
        \hline
        \hline
    \end{tabular}
            \caption{The accuracy results of NILM on different experimental settings. LF denotes low-frequency data and HF denotes high-frequency data.}
\vspace{-8mm}
\label{tab:nilm_results}
\end{table*}

\subsection{NILM Results}
We then study the application of SRP in Non-Intrusive Load Monitoring (NILM).
Appliance Load Monitoring (ALM) is essential for energy management solutions, allowing them to obtain appliance-specific energy consumption statistics that can further be used to devise load scheduling strategies for optimal energy utilization.
NILM is the process of identifying appliances and their working states using a single smart meter.
High-frequency data will increase the accuracy of NILM yet requires high-frequency smart meters which are more expensive.
We use the SRP results as an alternative to high-frequency data.

Three classic NILM algorithms are used for this study: KNN \cite{berges2010enhancing}, SVM \cite{wu2004distance} and Decision Tree \cite{rodriguez2004interval}.
For every algorithm and experiment setting, models are trained on both SRP results and high-frequency data and tested with SRP results.
For comparison, the models trained and tested on low-frequency data and high-frequency data are used as a baseline.
The accuracy results are shown in \tablename~\ref{tab:nilm_results}, which shows the reconstructed high-frequency data can lead to better appliance monitoring results without changing the monitoring appliances.
Overall, the SRP results of the experiment with $f_l=100\mathbf{Hz}$ and $\alpha=10$ have the best performance, which can increase the most accuracy among the three experiments.
It will be very helpful for the industry to reduce energy consumption with low-resolution data sensors.

\section{Discussion and Future Work}
\label{Sec:diss}
In this paper, we propose a novel machine learning problem - the SRP problem as reconstructing high-quality data from unsatisfactory sensor data in industrial systems.
SRP can also be applied to many important industrial fields.
State estimation (SE) is one of the most closely related scenarios.
SE focuses on estimating system states at a specific time point.
SRP will show good performances and play significant roles in the aforementioned area.

Industrial state estimation, as an important component of the industrial control process, is mainly performed to estimate system state variables under the state conditions characterized by a set of sensor measurements.
Since most of the large-scale industrial systems are deployed decades ago, their state estimations are performed mainly based on low-frequency sampling sensors due to the limitations of sensor and communication technologies.
However, the higher standards of modern industry on system efficiency and product quality require more accurate system state estimation.
Also, a new source of threats, cyber attackers, are bringing significant challenges to the security of industrial control processes \cite{liang2017review}.
For example, Stuxnet targets industrial supervisory control and data acquisition (SCADA) systems and programmable logic controllers (PLC) to cause substantial system damages\footnote{See \url{https://en.wikipedia.org/wiki/Stuxnet} for more information.}.
Both industrial upgrade and security defense call for more precise monitoring on system running states, while existing low-frequency sampling sensors cannot meet the requirement.

Take the power system state estimation as an example.
The power system state estimation is formulated as a non-convex optimization problem, which analyzes measurements of voltage and power at certain nodes to output complete system states (including power, voltage magnitude, and angle) \cite{monticelli2000electric}.
In the traditional power system, sensors (low-frequency) were deployed in the remote terminal units (RTUs), and some of the traditional sensors have been replaced by phasor measurement units (PMUs) (high-frequency) recently \cite{london2009redundancy}.
However, it is impossible to replace all traditional sensors with PMUs because PMU is extremely expensive.
Under such circumstances, on the one hand, low-frequency measurements cannot provide high-resolution observations on system states; on the other hand, lots of high-frequency measurements collected from PMUs are wasted because the industry has no idea how to co-utilize both the high-frequency data collected by a small number of PMUs and the low-frequency data collected by traditional RTUs.
This makes the power system state estimation a good application of SRP.
SRP can supplement enough super-resolution information in between two sets of measurements so as to percept system states more clearly.

It should be noticed that different from the environment with only a single sensor, state estimation relies on multiple sensors at a specific time point to estimate corresponding state variables, which relies on the domain knowledge of system internal relationships in the space domain (e.g., the relationships between the temperatures of different rooms in a large building).
Therefore, both the time domain and space domain information may be considered when applying SRP in industrial state estimation, which will be our future work. 

\bibliographystyle{ACM-Reference-Format}
\bibliography{sample-base}

\end{document}